\documentclass[aps,twocolumn,floats,prd,nofootinbib,showpacs,showkeys]{revtex4} % superscriptaddress,
\usepackage[dvips]{graphicx} % 
\usepackage{epsfig,amsmath}
\usepackage{amssymb}
\usepackage{rotate}
\usepackage{color}
\usepackage[ps2pdf=true]{hyperref}
\usepackage{bm}

\DeclareFontFamily{OT1}{pzc}{}
\DeclareFontShape{OT1}{pzc}{m}{it}%
             {<-> s * [1.00] pzcmi7t}{}
\DeclareMathAlphabet{\mathscr}{OT1}{pzc}%
                                 {m}{it}

\newcommand{\be}{\begin{equation}}
\newcommand{\ee}{\end{equation}}
\newcommand{\bea}{\begin{eqnarray}}
\newcommand{\eea}{\end{eqnarray}}
       
\newcommand{\refeq}[1]{Eq.~(\ref{eq:#1})}          
\newcommand{\refeqs}[2]{Eqs.~(\ref{eq:#1})--(\ref{eq:#2})}          
          
\newcommand{\reffig}[1]{Fig.~\ref{fig:#1}}          
\newcommand{\reftab}[1]{Tab.~\ref{tab:#1}}

\newcommand{\refsec}[1]{\S~\ref{sec:#1}}          
          
\newcommand{\EpiG}{8\pi G} %uniformity in spacing convention
 
\renewcommand{\v}[1]{\mathbf{#1}}
\newcommand{\ph}{\varphi}

\newcommand{\vn}{\bm{\nabla}}
\newcommand{\vx}{\v{x}}

\renewcommand{\l}{\lambda}

\newcommand{\eps}{\varepsilon}
\renewcommand{\d}{\delta}
\newcommand{\D}{\Delta}
\newcommand{\rhob}{\overline{\rho}}

\newcommand{\Mpch}{\,{\rm Mpc}/h}

\newcommand{\Lbox}{L_{\rm box}}

\newcommand{\Msunh}{\,M_{\odot}/h}

\newcommand{\Om}{\Omega_m}

\newcommand{\OL}{\Omega_\Lambda}

\renewcommand{\L}{\Lambda}

\newcommand{\Rv}{R_{\Delta}}
\newcommand{\Mv}{M_{\Delta}}
\newcommand{\Mdyn}{M_{\Delta,\rm dyn}}

\newcommand{\cv}{c_{\Delta}}
\newcommand{\Psiv}{\Psi_{\Delta}}

\newcommand{\s}{\sigma}

\newcommand{\Meff}{M_{\rm eff}}
\newcommand{\rscr}{r_{\rm scr}}

\newcommand{\g}{\mathscr{g}}
\newcommand{\gbar}{\overline{\g}}
\newcommand{\gvir}{\gbar_{\rm vir}}
\newcommand{\gvirf}[1]{\overline{\g}_{{\rm vir},#1}}
\newcommand{\fRbar}{\overline{f_R}}

\begin{document}

\title{Dynamical Masses in Modified Gravity}

\author{Fabian Schmidt}
\affiliation{Theoretical Astrophysics, California Institute of Technology M/C 350-17,
Pasadena, California  91125-0001, USA}

%Within certain limits,
%these models satisfy all current bounds from Solar System measurements
%and cosmology; in particular, they c
\begin{abstract}
Differences in masses inferred from dynamics, such as velocity dispersions or 
X-rays, and those inferred from lensing are a generic prediction of modified 
gravity theories.  Viable models however must include some non-linear 
mechanism to restore General Relativity (GR) in dense environments, which is 
necessary to pass Solar System constraints on precisely these deviations.  
In this paper, we study the dynamics within virialized structures in the 
context of two modified gravity models, $f(R)$ gravity and DGP.  The
non-linear mechanisms to restore GR, which $f(R)$ and DGP implement in very 
different ways, have a strong impact on the dynamics in bound objects; they
leave distinctive signatures in the dynamical mass-lensing mass relation
as a function of mass and radius.  We present measurements from 
N-body simulations of $f(R)$ and DGP, as well as 
semi-analytical models which match the simulation results to surprising
accuracy in both cases.  The semi-analytical models are useful for making the
connection to observations.  
Our results confirm that the environment- and scale-dependence of
the modified gravity effects have to be taken into account when
confronting gravity theories with observations of dynamics in galaxies and clusters.
\end{abstract}

\keywords{cosmology: theory; modified gravity; Dark Energy}
\pacs{95.30.Sf 95.36.+x 98.80.-k 98.80.Jk 04.50.Kd }

\date{\today}

\maketitle

%%%%%%%%%%%%%%%%%%%%%%%%%%%%%%%%%%%%%%%%%%%%%%%%%%%%%%%%%%%%%%%%%%%%%%%%
%%%%%%%%%%%%%%%%%%%%%%%%%%%%%%%%%%%%%%%%%%%%%%%%%%%%%%%%%%%%%%%%%%%%%%%%
\section{Introduction}
\label{sec:intro}

% acceleration, testing gravity
Gravity, as described by General Relativity (GR), is remarkably weakly
constrained in the present day on scales larger than a few AU.  Though 
measurements
from binary pulsar timing to the cosmic microwave background (CMB) and
big bang nucleosynthesis are all consistent with GR, there is still
room for order unity deviations in the cosmos today, on scales of kpc and larger.  
Thus, testing gravity on cosmological scales is an interesting frontier,
and the focus of much current research \cite{ZhangEtal,JainZhang,SongKoyama,KnoxSongTyson,Schmidt08,Song2006,JainZhang,Tsujikawa2008,ZhangfR,SongEtalDGP,Schmidt07,Uzan09}.
%Our knowledge of the initial conditions through the CMB and the
%relatively clean physics of gravitational collapse, as well as a rapidly
%growing observational data set are opening up a new window to testing
%gravity.

% Models, non-linear mechanisms
Any gravity theory that attempts to be complete has to satisfy stringent
Solar System constraints, and has to locally match the predictions of GR to 
within one part in $10^5$ there.  Only a few consistent models which modify gravity
appreciably on large scales, but restore GR locally are known.  Two of them will
be the subject of this study: $f(R)$ gravity \cite{Caretal03,NojOdi03,Capozziello:2003tk,Sotiriou:2008rp}, and the DGP
model \cite{DGP1}.  Within certain bounds placed by the CMB and
expansion history measurements in addition to Solar System tests, both 
theories can be made to satisfy all current constraints on gravity (including the
observation of an accelerating expansion).  In both models there exists a non-linear
mechanism to restore GR in high-density environments: the \textit{chameleon
effect} for $f(R)$, and the \textit{Vainshtein mechanism} for DGP.  Furthermore,
all currently known consistent modifications of gravity on large scales
include some variant of either of these mechanisms.  In
%\footnote{For example, TeVeS \cite{Bekenstein04}, though considerably more complicated as a whole, effectively invokes a
%chameleon mechanism to transition from the MOND limit to GR.}
order to be able to constrain these models with cosmological data, it is crucial
to correctly include the non-linear mechanisms.  Recently, N-body simulations
of $f(R)$ \cite{HPMpaper} and DGP \cite{DGPMpaper,ScII,DGPMpaperII} have been 
done which self-consistently solve the non-linear field equations together 
with the growth of structure (see also \cite{KW} for the first study of the
DGP case, using a different approach).   In principle, it has
become possible with these simulations to unlock the wealth of observations 
available on non-linear 
scales to probe gravity, albeit in a necessarily model-dependent way.

% dynamics in MG
It is well known that the additional degrees of freedom present
in modified gravity theories generically affect the {\it dynamical} potential,
which governs the propagation of non-relativistic bodies,
differently than the {\it lensing} potential, which governs the propagation of
massless particles such as light (e.g., \cite{BekSan94}).  
%A simple example are scalar-tensor
%theories:  since the principal effect of the scalar field is to conformally
%rescale the metric $g_{\mu\nu}\rightarrow e^{2\phi}\:g_{\mu\nu}$, and the 
%null geodesics light that travels on are not affected by a conformal rescaling 
%of the metric, gravitational lensing is in general not directly affected
%by the scalar field.  
Thus, comparing dynamical with lensing mass estimates is an interesting 
and quite generic probe of 
modifications to gravity.  In this paper, we study the signatures of $f(R)$ 
and DGP in dynamical observables such as velocity dispersions, compared
to lensing which measures essentially the ``true'' mass (i.e. the integral over the
rest-frame density) in both models.  

Constraints on the difference between dynamical and lensing potential are 
often phrased in terms of the post-Newtonian
parameter $\gamma_{\rm PPN}$ [\refeq{gamma} below], in analogy to Solar System tests.  In general, 
however, the departures from GR cannot be encapsulated by a single parameter
but are functions of scale, time and the local environment.  In particular,
this is the case for both $f(R)$ and DGP.  Hence, we introduce a more generally
applicable quantity $\g$ [\refeq{gdef}] which is defined directly via the modified forces, and is
well suited for predictions in the context of $f(R)$ and DGP as well as 
for constraints from observations.  

Velocities of extragalactic objects are measured through their redshifts $z$,
which receive a contribution $|\D z| = v_{\parallel}/c$ from the 
line-of-sight velocity $v_{\parallel}$.  
In the cosmological context, there are two regimes where the dynamics
of matter can be understood fairly easily: on very large scales,
linear perturbation theory in the matter density is valid, simplifying
the theoretial predictions.  Large-scale velocity fields can be measured 
through the redshift
distortion of the power spectrum, which thus offers a probe of the dynamical 
potential \cite{Sc04,ZhangLBD07}.  On small scales, most of 
the observable matter lies in gravitationally bound dark matter halos.  In 
this regime, for relaxed systems, the velocity distribution of 
collisionless objects such as dark matter, 
galaxies or stars is related to the dynamical potential by the virial theorem.
  For collisional particles
such as diffuse gas, this relation is given by
hydrostatic equilibrium.  The virial or thermal velocities can be observed
as velocity dispersion of stars in galaxies, galaxies in clusters, or as
X-ray or Sunyaev-Zeldovich signal from diffuse gas in clusters.  Also,
the redshift-space matter power spectrum on small scales is a probe
of virial velocities \cite{PD96,Sc04}.

This paper is concerned with the latter regime, and our goal is to study
the dynamics of matter in halos.  Since these are highly
non-linear systems, rigorous results can only be obtained via
N-body simulations.  We therefore present measurements from the
modified gravity simulations of $f(R)$ and DGP 
\cite{HPMpaper,DGPMpaper,DGPMpaperII}.  However, for many practical
purposes including comparison with observations, it is necessary to go
beyond the simulation results which have limited
resolution and cover only a few points in the parameter space of the models.  
Thus, a sufficiently accurate semi-analytic model of the dynamics in modified 
gravity is desirable to bridge the gap with observations.  

Fortunately, we can make some justified assumptions which simplify the
problem greatly: first, since we are concerned with sub-horizon scales, we
employ the quasi-static approximation, neglecting time derivatives and
assuming the halos are in steady state.  Further, we assume spherically
symmetric halos.  While certainly not realistic, deviations from spherical
symmetry are not expected to affect the results qualitatively.  Throughout, we 
will assume a Navarro-Frenk-White (NFW) 
\cite{NFW} profile, although all derivations can easily be generalized to
different profiles.  The problem is then reduced to finding the solution
of the field equations for a spherically symmetric mass, and calculating
the modified gravitational force.  The accuracy of this simplified model
can then be benchmarked with the simulation results.

The paper is structured as follows.  In \refsec{th}, we introduce
our main observable, the modified gravitational force strength, and present
the theoretical expectations and semi-analytic models for $f(R)$ and DGP.  
\refsec{sim} contains the simulation results and comparisons with the
theoretical models.  We then discuss the application to observations
in \refsec{obs}.  We conclude in \refsec{concl}. 

%%%%%%%%%%%%%%%%%%%%%%%%%%%%%%%%%%%%%%%%%%%%%%%%%%%%%%%%%%%%%%%%%%%%%%%%
%%%%%%%%%%%%%%%%%%%%%%%%%%%%%%%%%%%%%%%%%%%%%%%%%%%%%%%%%%%%%%%%%%%%%%%%
\section{Theoretical Expectations}
\label{sec:th}

In this section, we derive theoretical expectations for the modified
gravitational forces and virial quantities measured in the simulations 
in \refsec{sim} and connected to observations in \refsec{obs}.  Gravitational
forces are given by the gradient of the dynamical potential $\Psi$, defined
via the perturbed FRW metric in Newtonian gauge:
\be
ds^2 = -(1 + 2\Psi)dt^2 + a^2(t) (1 + 2\Phi)d\v{x}^2.
\ee
As a reference point, we consider General Relativity (GR) in the Newtonian 
limit, where the dynamical potential satisfies the Poisson equation
\be
\nabla^2 \Psi = \nabla^2\Psi_N = 4\pi G\,\d\rho,
\label{eq:Poisson}
\ee
where $\d\rho=\rho/\rhob$ is the total matter overdensity.  Assuming 
spherical symmetry, which we will throughout, we can define
a parameter $\g$:
\be
\g(r) \equiv \frac{d\Psi/dr}{d\Psi_N/dr},
\label{eq:gdef}
\ee
which quantifies the strength of the gravitational force in modified
gravity relative to that which would be measured in GR given the same
density field.  $\g=1$ corresponds to unmodified forces.  
Here we have suppressed the dependence of $\g$ on the scale factor $a$;  
unless otherwise stated, we will always assume $a=1$.  
  
In the models we consider, the lensing potential satisfies\footnote{In $f(R)$,
there are corrections of order $\overline{f_R}$, and 
$|\overline{f_R}| \leq |f_{R0}| \leq 10^{-4}$ for the models we
consider; this is negligible compared to the $\mathscr{O}(0.1)$ effects we will
discuss.}
\be
\Psi_- \equiv \frac{1}{2}(\Psi-\Phi) = \Psi_N.
\label{eq:Psim}
\ee
Hence, $\g$ can be probed for example by comparing dynamical to lensing mass estimates
of a given object.  Such comparisons in the Solar System \cite{Will} or
for distant galaxies \cite{SchwabEtal09} are often phrased in terms of the
post-Newtonian parameter $\gamma_{\rm PPN}$:
\be
\gamma_{\rm PPN} = -\frac{\Phi}{\Psi} = 2\frac{\Psi_-}{\Psi} - 1
\stackrel{\rm BD}{=} 2\g^{-1} - 1.
\label{eq:gamma}
\ee
The last equality relating $\gamma_{\rm PPN}$ to our $\g$ parameter 
(where we have used \refeq{Psim}) is 
only valid when the force modifications are \emph{scale-independent}, such as
in Brans-Dicke (BD) type scalar-tensor theories.  Note that the PPN
parameter is formally defined via the potentials, while our $\g$ parameter
is derived in terms of forces.  Only forces, or more generally derivatives
of the potentials $\Psi$, $\Psi_-$ are observable, and a specific solution
of the potentials (e.g., the Schwarzschild metric) is used to infer 
$\gamma_{\rm PPN}$.  However, in the models we consider $\g$ is generally
scale-dependent, i.e. the scalar degrees of freedom do not follow the same 
scaling with distance as the GR potentials.  Hence, it is advantageous to 
define a parameter based directly on the forces, rather than $\gamma_{\rm PPN}$
which is not immediately linked to observables.  
%Note further that in general 
%the force modifications cannot be captured by a single parameter,
%but depend on the distance $r$.

In many practical cases, one is interested in a
weighted average of $\g$ over an object or region of space,
\be
\gbar_w = \frac{\int r^2 w(r)\:\g(r)\:dr}{\int r^2 w(r)\:dr},
\label{eq:gbar}
\ee
where $w$ is a weighting function depending on the precise observable
considered (we will turn to this in \refsec{obs}).  The key point is
that given a prediction for $\g(r)$ we can estimate any such weighted
average (as long as spherical symmetry is a sufficiently good approximation).  
In the next section, we will introduce one such averaged force modification
which is relevant for comparison with simulations.  
We will then review the Newtonian potential and scaling relations for a dark 
matter halo with NFW profile, before studying the same case for $f(R)$ and 
DGP gravity.

%%%%%%%%%%%%%%%%%%%%%%%%%%%%%%%%%%%%%%%%%%%%%%%%%%%%%%%%%%%%%%%%%%%%%%%%
\subsection{Virial theorem and velocity dispersion}
\label{sec:vir}

For the comparison with our dark matter-only simulations, it 
is useful to consider a collisionless system in virial equilibrium.  In 
that case, the virial theorem states that
\bea
W &=& -2T,\quad\mbox{where}\label{eq:virth}\\
W &\equiv& -\int d^3\vx\,\rho(\vx)\,\vx\cdot\vn\Psi(\vx),\label{eq:W}\\
T &\equiv& \frac{1}{2}\int d^3\vx\,\rho(\vx)\,\sigma_{v,3D}^2(\vx),
\label{eq:T}
\eea
denote the trace of the potential energy tensor and potential energy, respectively.  
Here $\sigma_{v,3D}^2 = 3\sigma^2_{v,1D}$ is the three-dimensional velocity
dispersion (see \refeq{sigmav} for our practical definition in terms of
dark matter particles).  Since the virial theorem is derived from the collisionless
Boltzmann equation, and is thus a consequence of energy-momentum conservation,
it is unchanged in any metric theory of gravity, and hence also in the
models we consider.  The modification
enters through the modified relation between the potential $\Psi$ and the
matter distribution.  

Note that in the cosmological context, we are not dealing with strictly
isolated systems, so that \refeq{virth} does not hold precisely.  
Nevertheless, the validity of $W = \alpha\: T$ for simulated dark matter
halos has been shown to hold to high accuracy 
\cite{ShawEtal,CuestaEtal,EvrardEtal}.  Here, the constant $\alpha$ depends
on the mass and radius definition chosen for the halos.  

In the spherically symmetric case, we can use the definition of $\g$ 
[\refeq{gdef}] to relate the potential
energy tensor and kinetic energy in modified gravity to the Newtonian
values $W_N$, $T_N$:
\be
\frac{W_{\rm mod. gr}}{W_N} = \frac{T_{\rm mod. gr}}{T_N} = \gvir,
\label{eq:Emod}
\ee
where $\gvir$ is given by \refeq{gbar} with a weighting function
\be
w_{\rm vir}(r) = \rho(r)\:r \frac{d\Psi_N}{dr}.
\label{eq:wvir}
\ee
The gradient of the Newtonian potential appearing here is uniquely
determined by the density $\rho(r)$, assuming that external tidal fields are negligible.

%%%%%%%%%%%%%%%%%%%%%%%%%%%%%%%%%%%%%%%%%%%%%%%%%%%%%%%%%%%%%%%%%%%%%%%%
\subsection{Newtonian potential of a halo}

Let us consider the GR case first.  We can
integrate \refeq{Poisson} to obtain:
\bea
\frac{d\Psi_N}{dr} &=& \frac{G\d M(<r)}{r^2},\\
\d M(<r) &\equiv& 4\pi\int_0^r dr'\,r'^2\,\d\rho(r').
\eea
Note that $\d M$ is defined in terms of the enclosed overdensity $\d\rho$.    
Imposing the condition $\Psi_N(r\rightarrow\infty)=0$, we can integrate
again and obtain:
\be
\Psi_N(r) = -\int_r^{\infty}dr' \frac{G \d M(<r')}{r'^2}.
\label{eq:PsiN}
\ee
Let us now consider an NFW halo with mass $\Mv$, defined as the mass contained
within a radius $\Rv$ so that the average density
within $\Rv$ is $\rhob\:\D$ (note that $\D$ here is arbitrary and does not have
to correspond to a certain ``virial'' overdensity).  The NFW profile has been 
shown to be a good match even to the halos in modified gravity simulations 
\cite{HPMhalopaper,DGPMpaperII}.  We define the 
corresponding concentration as $\cv = \Rv/r_s$.  We will consider an 
untruncated profile here;  while this overestimates the exterior mass somewhat,
it is closer to the profiles measured in simulations than the other simple
choice, a truncated profile.  Then, the density profile is given by
\bea
\rho(r) &=& 4\rho_s\: f_{\rm NFW}(r/r_s),\\
f_{\rm NFW}(y) &=& \frac{1}{y (1+y)^2},
\eea
where $\rho_s=\rho(r_s)$ is chosen so that the mass within $\Rv$ is $\Mv$, 
and we have
\bea
\d M(< r) &=& \Mv\frac{F\left(\cv r/\Rv\right)}{F(\cv)}\left[ 1 - 
\frac{F(\cv)\: (r/\Rv)^3}{F(\cv r/\Rv)\:\D}\right]\quad\quad  \label{eq:dM}\\
F(y) &=& -\frac{y}{1+y} + \ln (1+y).\label{eq:F}
\eea
The correction in square brackets in \refeq{dM} is usually neglected since
it is smaller than $\D^{-1}$, and we will do so here as well in order to
simplify the analytical expressions.  From this, we get
\be
\frac{d\Psi_N}{dr} = \frac{\Psiv}{\Rv} \frac{\Rv^2}{r^2}\: \frac{F\left(\cv r/\Rv\right)}{F(\cv)},
\ee
where the potential scales with $\Psiv$ defined by
\be
\Psiv \equiv \frac{G\Mv}{\Rv}.
\ee
We can integrate to obtain the potential for an {\it isolated} NFW halo:
\bea
\Psi_N(r) &=& -\Psiv\:E\left(\frac{r}{\Rv},\cv\right),\label{eq:PsiNFW}\\
E(x,c) &\equiv& \frac{(1+c)\ln(1+c x)}{(1+c) x\ln( 1+c) - c x}.
\eea
$E(0,c)\sim 5-12$ (for $c\sim 4-30$) gives the central depth of the potential
well for an isolated NFW halo with concentration $c$ in units of $\Psiv$.  
Note that in reality, the depth of the potential well will depend
on the large scale environment, so that \refeq{PsiNFW} will only give a rough
scaling.  $\Psiv$ in turn is given by:
\bea
\Psiv &=& \left( G\,\Mv\,H_0\right)^{2/3}\left(\frac{1}{2}\Om \D\right)^{1/3}\\
&=& \left( \frac{\Mv}{6.26\times10^{22}\Msunh}\right)^{2/3}\left(\frac{1}{2}\Om \D\right)^{1/3}\\
&=& 1.79\times 10^{-5} \left( \frac{\Mv}{10^{15}\Msunh}\right)^{2/3},
\eea
where for the last equality we have assumed $\Om=0.25$ and $\D=200$.
Note the scaling of $\Psiv$ with $\Mv^{2/3}$, because $\Mv$ and $\Rv$
are linked through the fixed overdensity $\D$.  Throughout, unless otherwise
stated we use the concentration relation of \cite{BulEtal01}:
\be
c(M,z) = \frac{9}{1+z} \left(\frac{M}{M_*(z)}\right)^{-0.13}.
\label{eq:cM}
\ee
Here, $M_* \approx 3.2\times10^{12}\Msunh$ for our fiducial $\L$CDM cosmology.  
Recently, more accurate expressions for the concentration have been found
\cite{MaccioEtal,ZhaoEtal09}.  However, our results are not very sensitive
to the concentration, hence we deem \refeq{cM} sufficient.  At the
very highest masses $\Mv \gtrsim 10^{15}\Msunh$ however, \refeq{cM} underpredicts
the concentration significantly (e.g., \cite{ZhaoEtal03,ZhaoEtal09}).  As a simple remedy, we take
$c = \max\{4,c(M)\}$ in place of $c(M)$ from \refeq{cM}.  

Finally, the weighted $\g(r)$ quantifying the modification to the potential
and kinetic energy [\refeq{Emod}] can be written as
\be
\gvir = \frac{\int_0^1 dx\:x\: F(\cv x) f_{\rm NFW}(\cv x) \g(x\,\Rv)}
{\int_0^1 dx\:x\: F(\cv x) f_{\rm NFW}(\cv x)},\quad
\label{eq:gvir}
\ee
where $x=r/\Rv$.

%%%%%%%%%%%%%%%%%%%%%%%%%%%%%%%%%%%%%%%%%%%%%%%%%%%%%%%%%%%%%%%%%%%%%%%%%%%%%
%\subsection{$\bm{f(R)}$}
\subsection{\textbf{\textit{f(R)}}}
\label{sec:fR}

$f(R)$ gravity (see \cite{Sotiriou:2008rp} for a review) is a modified action 
theory where the Einstein-Hilbert
Lagrangian $R/16\pi G$ is replaced with $[R+f(R)]/16\pi G$.  Throughout
this section $R$ denotes the Ricci scalar.  
$f(R)$ models correspond to scalar-tensor theories, where the scalar degree
of freedom is given by $f_R\equiv df/dR$ and mediates the relation between density
and space-time curvature. In order for the theory to be stable under perturbations, it is necessary that $f_R < 0$ \cite{HuSawickifR}. 

In the smooth background, the scalar field assumes a value of $\fRbar \equiv f_R(\bar R)$,
where $\bar R \propto H^2$ is the scalar curvature of the background.    
%Since we work at $z=0$ throughout, it is useful to define 
%$f_{R0} = f_R(\bar R_0)$, the value of the $f_R$ field
%in the background today.  
In this paper, we only consider models with
$|\fRbar| \leq 10^{-4}$, and will thus drop higher order terms in the $f_R$
field which simplifies the expressions.  In the quasi-static regime, 
the $f_R$ field and 
the dynamical potential are then determined from the density field by the 
following coupled equations:
\bea
\nabla^2\d f_R &=& \frac{1}{3}\left [\d R(f_R)- 8\pi G \d\rho\right],\label{eq:fReom}\\
\nabla^2\Psi &=& \frac{16\pi G}{3}\d\rho - \frac{1}{6}\d R(f_R).\label{eq:fRpsi}
\eea
Here, $\d$ stands for perturbations from the background value:
$\d f_R\equiv f_R - \fRbar$ and $\d R \equiv R - \bar R$.  $R$ and $\d R$ 
are non-linear functions of the field $f_R$, hence \refeqs{fReom}{fRpsi} are 
difficult to solve in general.  However, there are
two limiting cases which can be solved easily.  

First, consider the
case where $\fRbar$ is much larger than typical potential wells in the 
universe.  In that case, 
$\d f_R$ sourced by the r.h.s. of \refeq{fReom} is always much less
than $\fRbar$, and we can linearize the $\d R$ term:
\be
\d R \approx \frac{1}{f_{RR}(\bar R)} \d f_R,
\ee
where $f_{RR} =d^2f/dR^2$.  
\refeq{fReom} then becomes an equation for a massive scalar field with
$m_{f_R}^{-2} \equiv \l_C^{2} = 3 f_{RR}(\bar R)$.  We call the inverse mass
$\l_C$ of the field in the background the Compton 
wavelength.  In this
limit, $\d R \ll \bar R$ on scales smaller than $\l_C$.  \refeq{fRpsi}
then tells us that $\Psi = 4/3\:\Psi_N$, i.e. gravitational forces
are increased by 4/3 within the range of the $f_R$ field given by $\l_C$.  

In the opposite limit, both terms on the r.h.s. of \refeq{fReom} are
much larger than the l.h.s. $\sim\fRbar/r^2$ on the scales of interest.  
Since the field perturbation
is limited in magnitude to be less than $|\fRbar|$, $\d f_R$ has to adjust 
itself so that the two terms on the r.h.s. cancel to a high degree, in other words
\be
\d R(f_R) \approx 8\pi G\d\rho.
\ee
Hence, the GR expression is restored, and \refeq{fRpsi} yields
$\Psi = \Psi_N$ accordingly.  This is called the chameleon regime \cite{khoury04a}.

In order to determine the transition between these two regimes, we consider
the solution for a spherically symmetric mass.  Formally, we can write
the solution for $\d f_R$ as:
\bea
\d f_R(r) &=& \frac{2}{3}\frac{G \d\Meff(<r)}{r},\\
\d\Meff(r) &=& 4\pi \int_0^r dr'\:r'^2 \d\rho_{\rm eff}(r'),\label{eq:Meff}\\
\d\rho_{\rm eff}(r) &=& \d\rho(r) - \frac{\d R(r)}{8\pi G}.
\label{eq:rho_eff}
\eea
With these definitions, the modified dynamical potential satisfies
\be
\nabla^2\Psi = 4\pi G\:\left(\d\rho + \frac{1}{3}\d\rho_{\rm eff}\right).
\label{eq:PsifR}
\ee
\refeqs{Meff}{rho_eff} state that $\d\Meff \leq \d M$.  If the perturbation $\d f_R$ is small for 
{\it all} $r$ (which in general is only true far away from the body), we can 
neglect the $\d R$ term in 
\refeq{rho_eff}.  Then, $\d\Meff = \d M$ and we have $|\d f_R(r)| = 2/3 |\Psi_N|$.  
However, the maximal value $\d f_R$ can achieve is $|\fRbar|$,
in which case the $f_R$ field becomes 0.  Thus, we arrive at the following
condition:
\be
|f_{R0}| \leq \frac{2}{3}\Psi_N.
\label{eq:thinshell}
\ee
If the value of $\Psi_N$ for the body is larger than this, the field {\it must}
enter the chameleon regime.  Then, $\d\rho_{\rm eff}$ is non-zero only outside of the
radius where \refeq{thinshell} is met.  $\d\Meff$ is thus given by the mass 
outside of this radius which can be thought of as forming a thin shell.  
For this reason, \refeq{thinshell} is also called the thin-shell criterion.  
Since cosmological potentials range from $10^{-6}-10^{-5}$, we expect that
the chameleon mechanism will operate for background field values $\lesssim 10^{-5}$.

This general picture holds for any viable functional form of $f(R)$.  However,
in order to evaluate the effect on the dynamics quantitatively and to compare
with the N-body simulations, we have to adopt a specific model.  The
functional form used in the simulations \cite{HPMpaper,HPMpaperII,HPMhalopaper}
is the one of \cite{HuSawickifR} with $n=1$, i.e.
\be
f(R) = -2 \Lambda \frac{R/R_c}{R/R_c+1},
\label{eq:fRex}
\ee
parametrized by the two constants $\Lambda$ and $R_c$.  If the
present-day background curvature $\bar R_0$ is much greater than $R_c$,
which will be the case for the $f(R)$ models considered here, we can
expand \refeq{fRex} to first order in $R_c/R$, and define a new
parameter $f_{R0}=f(\bar R_0)$ so that
\be
f(R) = -2 \Lambda - f_{R0}\frac{\bar R_0^2}{R}.
\label{eq:fRap}
\ee
The first term supplies an effective cosmological constant yielding accelerated
expansion of the background.  The second term, controlled by $f_{R0}\ll 1$
determines the departures from GR, and yields corrections to the background
expansion of order $f_{R0}$.  Since we will consider models with 
$|f_{R0}| \leq 10^{-4}$, the background expansion is essentially indistuingishable from
$\L$CDM.  Taking the derivative of \refeq{fRap}, we obtain
the relation between the scalar field and the local curvature at the present
day:
\be
f_R = f_{R0}\frac{\bar R_0^2}{R^2}.
\ee
Furthermore, the Compton wavelength $\l_C \propto \sqrt{f_{RR}}\propto R^{-3/2}$.  
% !!!!!!!!!!!!!!!!!!!!!!!!!!!
\begin{figure}[t!]
\centering
\includegraphics[width=0.49\textwidth]{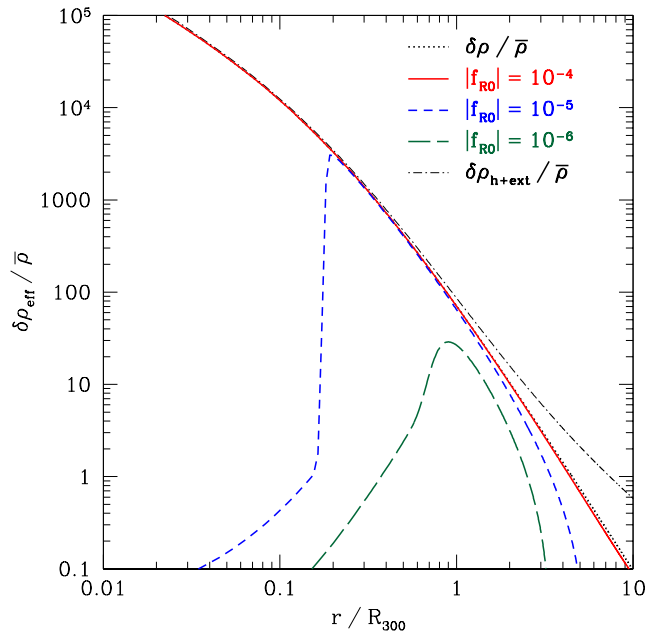}
\caption{$\d\rho_{\rm eff}$ [\refeq{rho_eff}] divided by the mean matter
density $\rhob$, determined from the numerical solution of the $f(R)$ field
equation for an NFW halo with $M_{300}=2\times 10^{14}\Msunh$ for different
values of $f_{R0}$.  Also shown is the matter density $\d\rho/\rhob$ of the
halo itself (dotted line almost matching the $10^{-4}$ field curve).  The
dash-dotted line shows a density profile which matches that measured in
simulations, including an additional external overdensity [\refeq{rhoext}].
\label{fig:rhoeff}}
\end{figure}
% !!!!!!!!!!!!!!!!!!!!!!!!!!!

Simulations were performed for a range of background field 
values $|f_{R0}|= 10^{-6}, 10^{-5}, 10^{-4}$.  From our discussion above,
we expect that the chameleon mechanism will operate in the intermediate
and small field cases, while it will be essentially absent for the
large field ($10^{-4}$).  In addition to the $f(R)$ simulations, ordinary
$\L$CDM simulations were performed using the same expansion history 
and initial conditions.  The cosmological parameters used in the simulations
are summarized in \reftab{param}.  

Given a density field such as that for an isolated NFW profile, one can
solve \refeq{fReom} numerically.  We have done so for the spherically
symmetric case using a one-dimensional relaxation algorithm (in fact we
solve for $u$ defined by $f_R = \exp(u)$ to avoid overshooting to $f_R>0$
\cite{HPMpaper}).  While only an approximation of the physical reality,
the spherically symmetric case allows for a much higher resolution
(at much smaller computing time) than achievable in the full 3D cosmological
simulations.  The boundary
conditions are given by $d f_R/dr = 0$ at $r=0$, and $\d f_R=0$ at
the outer edge of the grid, chosen here to be $r_{\rm max}=50\Mpch$.  
We use 4096 equally spaced grid points in $r$.  Once $\d f_R$ is known,
\refeqs{Meff}{PsifR} can be evaluated using $\d R(\d f_R)$, and 
the modified forces are given by
\be
\g_{f(R)}(r) = 1 + \frac{1}{3}\frac{\Meff(<r)}{M(<r)}.
\ee
\reffig{rhoeff} shows the ``effective density'' $\d\rho_{\rm eff}$
which sources the perturbation $\d f_R$ to the field, for a halo of mass
$2\times 10^{14}\Msunh$ and different values of $f_{R0}$.  For large values
of $|f_{R0}| \gtrsim 2\times 10^{-5}$, the thin shell condition is never
met, so that $\d\rho_{\rm eff} = \d\rho$ everywhere (except at very large $r$
where the field decays due to its finite $\l_C$).   For smaller
field values, we can see that a ``thin shell'' develops.  For
$|f_{R0}|=10^{-5}$ it is quite broad, while it narrows considerably
for a small field of $|f_{R0}|=10^{-6}$.  Note that the transition
to $\rho_{\rm eff}=0$ within the shell is very sharp, owing to the 
much smaller Compton wavelength within the body (recall $\l_C\propto R^{-3/2}$).

% !!!!!!!!!!!!!!!!!!!!!!!!!!!
\begin{figure}[t!]
\centering
\includegraphics[width=0.49\textwidth]{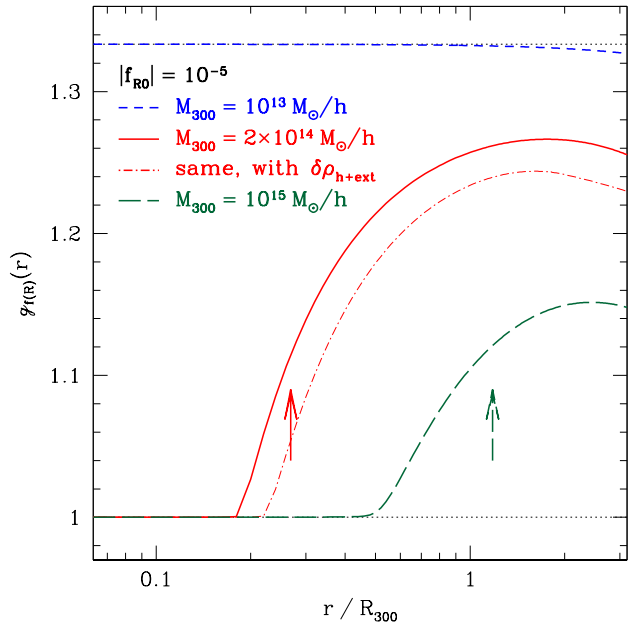}
\caption{$\g_{f(R)}$ as a function of the scaled radius $r/R_{300}$ for 
$|f_{R0}|=10^{-5}$ and different halo masses, from the numerical spherically
symmetric solution.  The low-mass halo is unscreeened,
showing the 4/3 force enhancement throughout, 
while higher mass halos are partially screened.  The arrows for the two more
massive halos indicate at which $r$ the condition \refeq{thinshell} is first
met.  For the $2\times 10^{14}\Msunh$ halo, we also show $\g_{f(R)}$ including
an external density field (dash-dotted line; see \reffig{rhoeff} and \refeq{rhoext}).
\label{fig:gfR}}
\end{figure}
% !!!!!!!!!!!!!!!!!!!!!!!!!!!
% !!!!!!!!!!!!!!!!!!!!!!!!!!!
\begin{figure}[t!]
\centering
\includegraphics[width=0.49\textwidth]{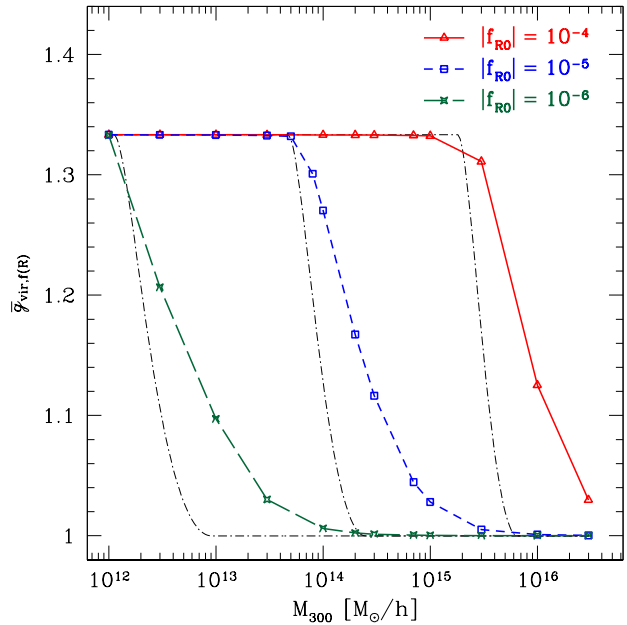}
\caption{Averaged modified gravitational force $\bar\g_{{\rm vir},f(R)}$ for
NFW halos as a function of halo mass for different values of $|f_{R0}|$.  The
points and thick lines show the numerical results from the 1D relaxation code.  
The thin black lines show the predictions of a simplified model \refeq{gvirmodel}.  
The behavior with mass is similar in the different models, with the transition
mass between unscreened and screened regimes shifting as expected by simple
estimates.
\label{fig:gvirfR}}
\end{figure}
% !!!!!!!!!!!!!!!!!!!!!!!!!!!

\reffig{gfR} shows $\g_{f(R)}(r)$ for $|f_{R0}|=10^{-5}$ and different
halo masses.  The $10^{13}\Msunh$ halo is unscreeened, showing the 4/3 force 
enhancement throughout. The $2\times 10^{14}\Msunh$ halo is partially screened,
while the $10^{15}\Msunh$ halo is screened to a large extent within
$R_{300}$.  For the latter two cases, we also indicate the screening radius $\rscr$
where, going from the outside in, the thin shell condition \refeq{thinshell} 
is first met.  This radius serves as an 
indication of whether a given mass is screened, and roughly to what extent.  
As \refeq{thinshell} shows, the screening radius depends on the depth
of the potential well, which is influenced by the large-scale environment.  
To investigate this effect, we have added an external large-scale density field to 
the NFW profile, roughly matched to the halo profiles in our simulations
\cite{HPMhalopaper} at large radii:
\be
\frac{\d\rho_{\rm h+ext}(r)}{\rhob} = \max\left\{ \frac{\d\rho_{\rm NFW}}{\rhob},\;\;
30 \left(\frac{r}{\Rv}\right)^{-1.46}\right\},
\label{eq:rhoext}
\ee
where $\d\rho_{\rm NFW}$ is the halo overdensity given by the NFW profile,
and the external density is smoothly cut off at $\sim 40\Mpch$.  
$\d\rho_{\rm h+ext}$ is shown as dash-dotted
line in \reffig{rhoeff}.  The resulting $\g_{f(R)}$ including the external 
density field is shown in \reffig{gfR} for the intermediate mass halo.  
Clearly, the field is screened at somewhat larger radii in this case, and 
$\g_{f(R)}$ is smaller than
that predicted for the NFW profile alone by about 0.04 in the transition
region.  Since halos can reside in a variety of environments, we expect
significant scatter in the strength of the modified forces within
halos in the $f(R)$ case, halos in overdense regions being screeend more
strongly than those in average or underdense regions.  Further, we expect that
the environment-dependence will be more significant for lower mass halos
than for massive halos ($\gtrsim 10^{14}\Msunh$), since the former can
be affected by a massive halo nearby, while the latter usually dominate
their environment.  We also investigated the effect of varying
the halo concentration by $\pm 20$\%; the impact on $\g_{f(R)}$ is small in 
comparison with the effects of the large scale environment however.  

Finally, using the results for $\g_{f(R)}(r)$ we can evaluate \refeq{gvir}.
\reffig{gvirfR} shows $\g_{{\rm vir},f(R)}$ as a function of mass for different 
values of the background field $f_{R0}$.  The thick lines and points show
the numerical results from the relaxation code.  For the strongest field, 
only the most massive
halos (more massive than found in our limited volume simulations) are
chameleon-screened.  For the weakest field, all halos above 
$M\sim 10^{13}\Msunh$ are expected to be screened, while for the intermediate
field the transition scale is around $10^{14.5}\Msunh$, relevant
for galaxy clusters.  We will compare the predictions for both $\g(r)$ and $\gvir$ with 
simulation results in \refsec{sim}.

As a simple analytic model for the numerical results, we make the assumption
that all mass of the halo outside of $\rscr$ contributes to $\d\Meff$.  This
results in the following simple prescription:
\bea
\g_{f(R)}(r) &\approx& 1 + \frac{1}{3}\frac{M(<r) - M(<\rscr)}{M(<r)}\nonumber\\
&=& 1 + \frac{1}{3}\left(1 - \frac{F(\cv \rscr/\Rv)}{F(\cv r/\Rv)}\right).
\label{eq:gvirmodel}
\eea
We then form the same average via \refeq{gvir}.  As shown in 
 \reffig{gvirfR} (thin black lines), this approximation predicts the onset
of the chameleon screening quite well, though the predicted transition between unscreened
and screened regimes is somewhat too sharp.  Nevertheless,
this simple model can be useful in interpolating the numerical results for
different values of $f_{R0}$.  

%%%%%%%%%%%%%%%%%%%%%%%%%%%%%%%%%%%%%%%%%%%%%%%%%%%%%%%%%%%%%%%%%%%%%%%%%%
\begin{table}[b!]
\caption{Parameters of the simulated $f(R)$ and DGP cosmologies.  For each
model, GR simulations with identical expansion history and initial conditions
were also performed.\label{tab:param}} 
\begin{center}
  \leavevmode
  \begin{tabular}{l||l||l|l|l}
\hline
  & $f(R)$ & sDGP & nDGP--1 & nDGP--2 \\
%$10^{-4},\:10^{-5},\:10^{-6}$
\hline
$\Om$ & \  0.24 & \  0.258 & \  0.259 & \  0.259 \\
$\OL$ (eff.) & \ 0.76 & \ 0 & \ 0.741 & \ 0.741 \\
$\lg |f_{R0}|$ & --4, --5, --6\  & \  \  - & \  \  - & \  \  - \\
$r_c$~[Mpc]& \  \  - & \ 6118 & \  500 & \  3000 \\
$\beta(a=1)$ & \  \  - & \  -1.15 & \  1.21  & \  2.25 \\
$H_0$~[km/s/Mpc] & \  73.0 & \  66.0 & \  71.6 & \  71.6 \\
\hline
$100\,\Omega_b\,h^2$ & \  2.23 & \  2.37 & \  2.26 & \  2.26 \\
$n_s$ & \  0.958 & \  0.998 & \  0.959 & \  0.959 \\
$10^9\:A_s(0.05\,{\rm Mpc}^{-1})$ & \  2.24 & \  2.02 & \  
2.11 & \  2.11 \\
\hline\hline
$\sigma_8(\Lambda\rm CDM)$\footnote{Linear power spectrum normalization today
of a $\Lambda$CDM model with the same primordial normalization.}
  & \  0.796 & \  0.657 & \  0.789 & \  0.789 \\
\hline
\end{tabular}
\end{center}
\end{table}
%%%%%%%%%%%%%%%%%%%%%%%%%%%%%%%%%%%%%%%%%%%%%%%%%%%%%%%%%%%%

%%%%%%%%%%%%%%%%%%%%%%%%%%%%%%%%%%%%%%%%%%%%%%%%%%%%%%%%%%%%%%%%%%%%%%%%%%%%%
\subsection{DGP}
\label{sec:DGP}

In the DGP braneworld scenario \cite{DGP1}, matter and radiation live
on a four-dimensional brane in five-dimensional Minkowski space.  The action
is constructed so that on scales larger than the \textit{crossover scale}
$r_c$, gravity is five-dimensional, while it becomes four-dimensional
on scales smaller than $r_c$.  This model admits a homogeneous cosmological
solution on the brane which obeys a modifed Friedmann equation \cite{Deffayet01}:
\be
H^2 \pm \frac{H}{r_c} = 8\pi G\: [\rhob + \rho_{\rm DE}].
\ee
The sign on the l.h.s. is determined by the choice of embedding of the brane.  
The negative sign is called the \textit{self-accelerating} branch, since
it allows for accelerated expansion even in the absence
of a cosmological constant.  The positive sign is called the \textit{normal}
branch, which does not exhibit self-acceleration.  Here, we consider
models of both branches (see \cite{DGPMpaper,DGPMpaperII,DGPhalopaper}):  
a self-accelerating model without a $\L$ term ($\rho_{\rm DE}=0$), {\it sDGP}, 
where $r_c\sim 6000$~Mpc is
adjusted to best match CMB and expansion history constraints \cite{FangEtal}
(note that this model is in $\sim 4-5\sigma$ conflict with current data);  
and normal-branch models with a dark energy component $\rho_{\rm DE}$ adjusted 
so that the expansion history is exactly $\L$CDM \cite{DGPMpaperII}.  In that
case, $r_c$ is a free parameter, and we chose values of 500~Mpc ({\it nDGP--1})
and 3000~Mpc ({\it nDGP--2}).  The remaining cosmological parameters are
summarized in \reftab{param}.  For both sDGP and nDGP models, we have also performed
ordinary GR simulations employing the same expansion history and initial 
conditions as for the DGP simulations.

On sub-horizon scales, and scales smaller than the crossover scale $r_c$,
DGP braneworld models can be accurately described
as scalar-tensor theory \cite{NicRat}, where the brane-bending mode $\ph$ mediates
an additional attractive (normal branch) or repulsive (self-accelerating branch)
force.  Gravitational forces in DGP are governed by:
\be
\vn\Psi = \vn\Psi_N + \frac{1}{2}\vn\ph.
\label{eq:PsiDGP}
\ee
The $\ph$ field is sourced by matter overdensities similarly to the usual
GR potentials, but has quadratic self-interactions which suppress
the field once density contrasts become non-linear.  The full equation
for the $\ph$ field is (assuming $a=1$; see e.g. \cite{KoyamaSilva}):
\be
\nabla^2 \ph + \frac{r_c^2}{3\beta} [ (\nabla^2\ph)^2
- (\nabla_i\nabla_j\ph)(\nabla^i\nabla^j\ph) ] = \frac{\EpiG}{3\beta} \delta\rho .
\label{eq:phiQS}
\ee
Here $\beta$ is determined by the expansion rate $H(a)$ via
\be
\beta(a) = 1 \pm 2 H(a)\, r_c \left ( 1 + \frac{\dot H(a)}{3 H^2(a)} \right ),
\ee
where the positive (negative) sign are valid for the normal (self-accelerating)
branch.  
The present-day values for $\beta$ are given in \reftab{param}.  

While analytical solutions to \refeq{phiQS} do not exist in the general case, 
the case of a spherically symmetric
mass is solvable in terms of closed expressions \cite{LueEtal04,KoyamaSilva}.  In particular, one 
obtains the following equation for the gradient of $\ph$ \cite{DGPhalopaper}:
\bea
\frac{d\ph}{dr} &=& \frac{G \d M(<r)}{r^2} \frac{4}{3\beta} g\left(\frac{r}{r_*(r)}\right)\label{eq:dphdr}\\
g(\xi) &=& \xi^3 \left(\sqrt{1+\xi^{-3}} - 1\right).
\eea
$r_*(r)$ 
in \refeq{dphdr} is the $r$-dependent \textit{Vainshtein radius} defined
as:
\be
r_*(r) = \left(\frac{16 G \d M(<r) r_c^2}{9\beta^2}\right)^{1/3}.
\ee
Note that $r/r_*$ is a function of the average overdensity $\d\rho(<r)$ 
within $r$.  Specifically, scaling to a halo with mass $\Mv$ and radius
$\Rv$ determined by a fixed overdensity $\D$ and neglecting the small
difference between $M$ and $\d M$, we have:
\be
\frac{r}{r_*(r)} = (\eps\: \D)^{-1/3}\: x \left(\frac{M(<x)}{\Mv}\right)^{-1/3},
\label{eq:yofx}
\ee
where $x = r/\Rv$ and the quantity $\eps$ is determined by the
background cosmology:
\be
\eps = \frac{8}{9\beta^2} (H_0 r_c)^2 \Om a^{-3}.
\ee
At $a=1$, $\eps = 0.32$ for sDGP, and 0.002/0.023 for nDGP--1/nDGP--2, respectively.  
Using \refeq{gdef},~(\ref{eq:PsiDGP}),~(\ref{eq:dphdr}), we then have
\be
\g_{\rm DGP}(r)=  1 + \frac{2}{3\beta}\: g\left(\frac{r}{r_*(r)}\right).
\label{eq:gDGP}
\ee
On large scales where $\d\rho(<r)\ll \rhob$,  $r$ is much larger than $r_*$ 
so that $g(\xi)\rightarrow 1/2$ and $d\ph/dr$ becomes simply proportional to
$d\Psi_N/dr$.  In this limit, $\g_{\rm DGP}=\g_{\rm DGP,lin} = 1 + 1/(3\beta)$.  This is the same expression one would obtain by simply neglecting the
non-linear terms in \refeq{phiQS}.  
  On small scales where 
$r \ll r_*$, modified forces are suppressed by $(\eps\bar\d)^{-1/2}$,
where $\bar\d=\d\rho(<r)/\rhob$ is the average overdensity within $r$.  

Note that the specific tensorial structure of the non-linearities in
\refeq{phiQS} is crucial to recover the linearized expresion 
$\g_{\rm DGP,lin}$.   It is possible to simplify \refeq{phiQS} by 
neglecting the tensorial structure, resulting in a Poisson equation for
$\ph$ with a source term given by a non-linear function of $\d\rho$ \cite{KW}.
However, this simplification qualitatively changes
the large-distance behavior of the Vainshtein mechanism \cite{KW,DGPhalopaper}.  
Thus, it will turn out to be crucial that the simulations solve the full
\refeq{phiQS} for our comparison with the theoretical predictions
from \refeq{dphdr} and \refeq{gDGP}.

% !!!!!!!!!!!!!!!!!!!!!!!!!!!
\begin{figure}[t!]
\centering
\includegraphics[width=0.49\textwidth]{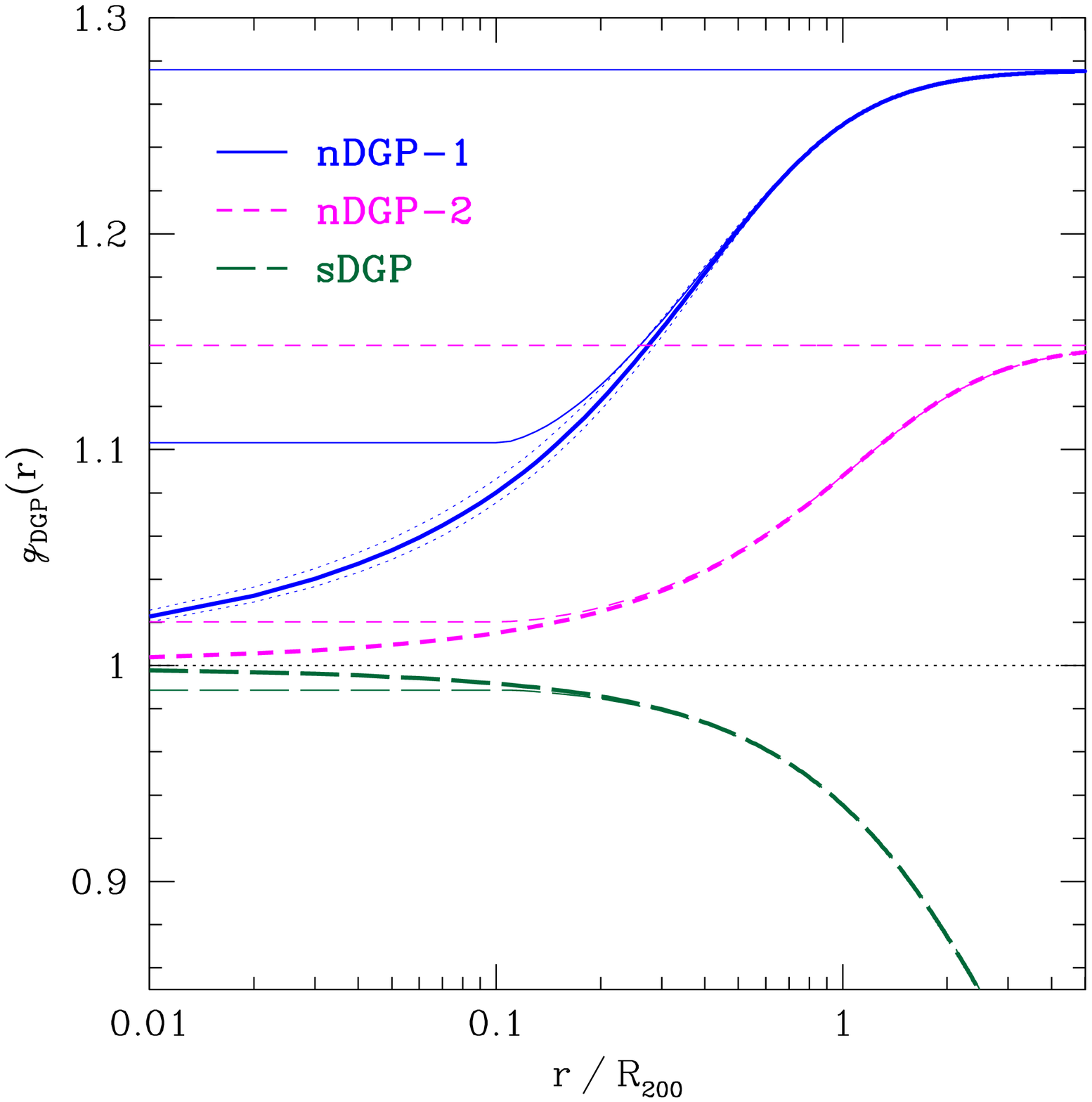}
\caption{$\g_{\rm DGP}(r)$ [\refeq{gDGP}] as function of the scaled radius
$r/\Rv$ for an NFW halo, for the DGP models.  The thin horizontal lines show the linearized
expression $\g_{\rm DGP,lin}=1+1/(3\beta)$, while the thin lines deviating at
small $r$ show the results when using a ``capped'' density profile 
with $r_{\rm cap}=0.125\Mpch$ (see \refsec{DGPsim}).  In all cases, we 
assumed $\D=200$ and a concentration of $\cv=5$.  For nDGP--1, we also show
the effect of varying the concentration by $\pm 20$\% (dotted lines).  Note that $\g_{\rm DGP}$ is indepedent of the halo mass.   
\label{fig:gDGP}}
\end{figure}
% !!!!!!!!!!!!!!!!!!!!!!!!!!!

% !!!!!!!!!!!!!!!!!!!!!!!!!!!
\begin{figure}[t!]
\centering
\includegraphics[width=0.49\textwidth]{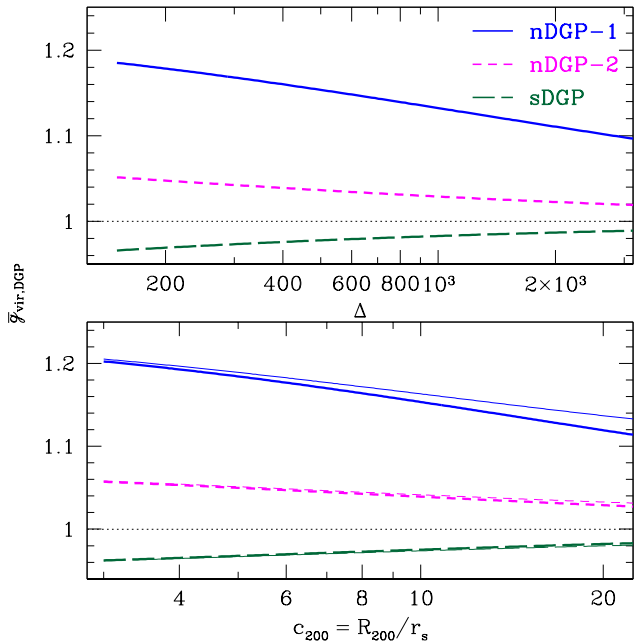}
\caption{Averaged force deviation $\gvir$ [\refeq{gvir}] for DGP models
as function of the overdensity $\D$ (\textit{top panel}) and the
halo concentration $c_{200}$ for an NFW halo (\textit{bottom panel}).  
For the top panel we have scaled the concentration with $\Rv$ to keep $r_s$ 
fixed (corresponding to $c_{200}\approx 5.8$).  
The thin lines in the bottom panel show 
the results for a density profile capped at $r_{\rm cap}=0.125\Mpch$ 
for comparison with simulation results (assuming $M_{200}=10^{14}\Msunh$; see \refsec{DGPsim}).
\label{fig:gvirDGP}}
\end{figure}
% !!!!!!!!!!!!!!!!!!!!!!!!!!!

Note that in the Vainshtein limit, 
\be
\ph(r\ll r_*) \approx C \frac{G \d M(<r)}{r_*} \propto \Psi_N(r) \frac{r_*}{r},
\ee
where $C$ is a constant of order unity \cite{DGPhalopaper}.  
Hence, the $\ph$ field itself is suppressed less than the modified forces by 
a factor of $(r/r_*)^{1/2}$.  However, only the forces
are observable.  This shows that in theories with non-linear interactions
such as DGP, quantifying departures from GR in terms of forces is more 
appropriate than the parameter $\gamma_{\rm PPN}$ defined in terms of the 
potentials [\refeq{gamma}].  

For a mass profile with constant density (``tophat''), the force
enhancement in \refeq{gDGP} is independent of radius; for more
general profiles however, this is not the case (see also \cite{DGPhalopaper}
for a detailed discussion).  \reffig{gDGP} shows the relative force enhancement $\g(r)$
as function of radius $r/\Rv$ in the case of an NFW halo, for the
different DGP models (\cite{DGPMpaper,DGPMpaperII}, \reftab{param}).  We also
show the ($r$-independent) linearized value $\g_{\rm DGP,lin}$ for the nDGP
models which is recovered only
at very large scales when the average density becomes $\lesssim \eps^{-1}$ 
(for sDGP, $\g_{\rm DGP,lin}\approx 0.76$ is beyond the range of the plot).  

  Since $\g_{\rm DGP}$ only depends on the average overdensity $\d\rho/\rhob$,
which is completely determined by $\D$ and $\cv$, the force enhancement does
not directly depend on the halo mass.  Also, it is insensitive to the 
large-scale environment
of the halo.  These are two crucial distinctions from the $f(R)$ case.

\reffig{gvirDGP} shows $\gbar_{\rm vir,DGP}$ defined in \refeq{gvir} as
a function of the overdensity $\D$ (keeping $r_s$ fixed at a value expected
for a $10^{14}\Msunh$ halo), and the halo 
concentration.  Clearly, $\gbar_{\rm vir,DGP}$ does depend somewhat on the halo 
profile and the overdensity criterion chosen.  The general trend is
that more concentrated halos lead to a stronger suppression of the modified
forces, since they have higher average density at small radii.   The same 
holds when increasing $\D$.  The dependence on $c$ and $\D$
is strongest for nDGP--1 which also shows the strongest evolution of 
$\g_{\rm DGP}(r)$.  The dependence on the density profile has to be taken 
into account when comparing with 
simulation results (\refsec{DGPsim}), as well as for the comparison with
observations which measure the dynamical mass within different $\Rv$ (\refsec{obs}).  
  
%%%%%%%%%%%%%%%%%%%%%%%%%%%%%%%%%%%%%%%%%%%%%%%%%%%%%%%%%%%%%%%%%%%%%%%%%%%
\begin{table}[b!]
\caption{Number of runs for each box size and minimum mass cuts for $\s_v^2$
and $\gvir$ measurement.} 
\begin{center}
  \leavevmode
  \begin{tabular}{c|c|c c c c}
  && \multicolumn{3}{|c}{$L_{\rm box}$ ($h^{-1}$ Mpc)} \\ %\vspace{1mm}\\
  \cline{3-6} 
  
& Model &\ \ $400$\ \ & $256 $\ \ \  & $128$\ \ \  & $64 $\ \ \  \\
\hline
\# of\ \ & $f(R)$\footnote{For each value of $|f_{R0}|$.} & 6 & 6 & 6 & 6\\
runs \ \ & sDGP & 6 & 6 & 6 & 6\\
     & nDGP\footnote{For nDGP--1 and nDGP--2 each.} & 3 & 3 & 3 & 6\\
\hline
\multicolumn{2}{c|}{$M_{\rm h, min}$ ($10^{14} \Msunh$)\ \ } & 63.5 & 16.7 & 2.08 & 0.26 \\
\hline
%\multicolumn{2}{c|}{$k_{\rm fun}=\pi/L_{\rm box}$  ($h$ Mpc$^{-1}$)\ \            } & 0.008 & 0.012 & 0.025 & 0.049 \\
%\hline
\multicolumn{2}{c|}{$r_{\rm cell}$ ($h^{-1}$ Mpc)\ \            } & 0.78 & 0.50 & 0.25 & 0.125 \\
\hline
\end{tabular}
\end{center}
\label{tab:boxes}
\end{table}
%%%%%%%%%%%%%%%%%%%%%%%%%%%%%%%%%%%%%%%%%%%%%%%%%%%%%%%%%%%%%%%%%%%%%%%%%%%

%%%%%%%%%%%%%%%%%%%%%%%%%%%%%%%%%%%%%%%%%%%%%%%%%%%%%%%%%%%%%%%%%%%%%%%%
%%%%%%%%%%%%%%%%%%%%%%%%%%%%%%%%%%%%%%%%%%%%%%%%%%%%%%%%%%%%%%%%%%%%%%%%
\section{Comparison with Simulations}
\label{sec:sim}

In order to benchmark our theoretical expectations, we will now
compare them to the results of the self-consistent N-body simulations 
of $f(R)$ gravity presented in \cite{HPMpaper,HPMpaperII} and 
of DGP \cite{DGPMpaper,DGPMpaperII}.  For each model, we have
simulated several box sizes.  The number of runs for each model and
box size, as well as the grid resolution are summarized in \reftab{boxes}. 
Halos are identified using a spherical overdensity halo finder as described
in \cite{HPMhalopaper,DGPhalopaper}.  The halo finder returns the 
center-of-mass position as well as mass $\Mv$ of the halo as determined from 
the particles within $\Rv$, such that $\Mv/(4\pi/3 \Rv^3) = \rhob \D$.  
Our choice of $\D$ is the one adopted in \cite{HPMhalopaper,DGPhalopaper}:
$\D=300$ for the $f(R)$ simulations, and $\D=200$ in the DGP case.  
Our particle-mesh simulations are of limited resolution, and we can only
use the best-resolved halos for our study, i.e. massive halos in the
two smallest boxes.  This limits our statistical sample of halos.  

First, in order to measure $\g(r)$ as function of radius, we select the
most well-resolved halos whose radii $\Rv$ are at least 10 grid cells, which
is only satisfied for halos in our smallest box, $\Lbox=64\Mpch$.  For this
box, this corresponds to a minimum mass of $\sim 1.6\times 10^{14}\Msunh$,
which depending on the model results in a very small sample of 2--40 halos.  
For each halo, we then measure contributions to the potential energy
$W(r)$ in spherical shells around the center-of-mass via
\be
W(r) = \frac{1}{N}\sum_{|r_i-r|\leq\D r} (\vx_i - \vx_h)\cdot\vn\Psi(\vx_i),
\label{eq:Wofr}
\ee
where the sum runs over particles whose distance $r_i = |\vx_i-\vx_h|$ 
from the center-of-mass of the halo is within the radial bin, and $N$ is
the number of contributing particles.  The derivative
of the potential is evaluated at the position of each particle in the
same way as it is done in the particle propagation of the N-body simulation 
(bilinear interpolation).  In addition to $W(r)$ derived from the dynamical 
potential $\Psi$, we also measure the corresponding Newtonian quantity $W_N(r)$,
where the Newtonian potential $\Psi_N$ is determined from \refeq{PsiN} using the
same density field.  The ratio of the two is our estimated force
enhancement:
\be
\g_{\rm meas}(r) = \frac{W(r)}{W_N(r)}.
\label{eq:gsim}
\ee
To some extent, resolution effects can be expected to cancel out in
\refeq{gsim}.  Below we will show profiles down to $r = r_{\rm cell}$,
though one should keep in mind that the $\g$ profiles cannot be considered
reliable below $r \sim 4 r_{\rm cell}$.  

Due to the resolution requirements and small sample size, we cannot study 
any evolution with mass in the $\g(r)$ profiles.  The stringent resolution 
requirements can be relaxed somewhat if we only measure an average
force enhancement, for example $\gvir$.  Assuming a scaling following
the virial theorem, we can either measure 
an average of $\vx\cdot\vn\Psi$, related to the
trace of the potential energy tensor $W$ (\refeq{virth}); or we can
measure the velocity dispersion, related to the
kinetic energy given by \refeq{T}.  The first approach has the advantage
that we can measure $\gvir$ on a halo-by-halo basis, by calculating
$W$ using both the modified potential $\Psi$ and the Newtonian potential $\Psi_N$ 
(similarly to what was done for $\g_{\rm meas}(r)$).  The estimator
of $\gvir$ for a given halo is then defined by:
\be
\gvirf{\rm meas} = \frac{\sum_{r_i < \Rv} (\vx_i - \vx_h)\cdot\vn\Psi(\vx_i)}
{\sum_{r_i < \Rv} (\vx_i - \vx_h)\cdot\vn\Psi_N(\vx_i)},
\label{eq:gvirsim}
\ee
where the sum runs over particles within the halo radius $\Rv$.  Note that
the sum over particles automatically results in a density weighting of $\g$.  
Again, we expect that in this ratio resolution issues cancel to a certain
extent.  Some effects of the finite resolution will become apparent when
comparing with the theoretical predictions below.

% !!!!!!!!!!!!!!!!!!!!!!!!!!!
\begin{figure}[t!]
\centering
\includegraphics[width=0.49\textwidth]{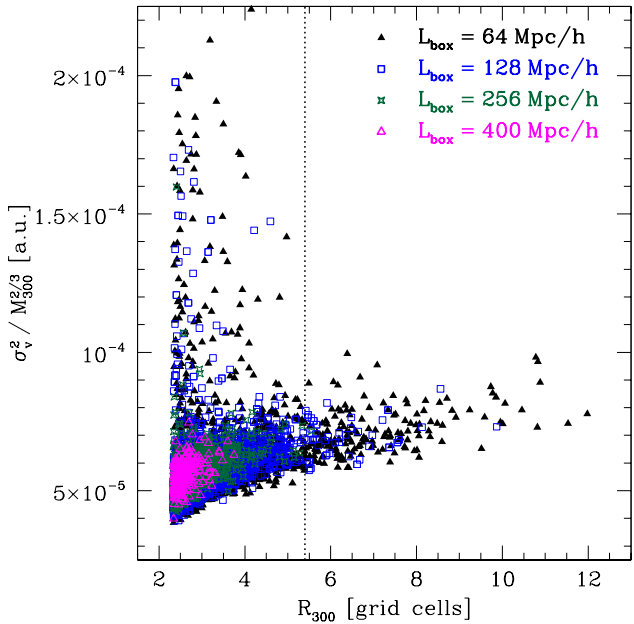}
\caption{Velocity dispersion $\sigma_v$ scaled to the virial expectation
($\sigma_v^2 \propto M^{2/3}$), measured for halos in the GR simulations,
as a function of the halo radius in grid cells.  Velocity 
dispersions are only reliably measured for the most well-resolved halos with
$R_{300} \geq 5.4$~grid cells (indicated by the vertical line). 
\label{fig:sigmav_res}}
\end{figure}
% !!!!!!!!!!!!!!!!!!!!!!!!!!!

The second approach, measuring the halo velocity dispersions is also interesting
since it gives an independent estimate of the modified forces.  Specifically, 
we define the (one-dimensional) velocity dispersion of particles in a halo
as follows:
\bea
\sigma_v^2 &=& \frac{1}{3 N_p} \sum_{|\vx-\vx_h|<R_\D} (\v{v}_i-\v{v}_h)^2,
\label{eq:sigmav}\\
\v{v}_h &=& \frac{1}{N_p} \sum_{|\vx-\vx_h|<R_\D} \v{v}_i,
\eea
where the sum runs over particles within the halo radius $\Rv$, $N_p$ is 
the number of those particles and 
$\v{v}_i-\v{v}_h$ denotes the velocity of the particle with respect to 
the center-of-mass of the halo.  
Note that in our normalization of $\sigma_v$, the kinetic energy
\refeq{T} is given by $T = 3/2\:\Mv \sigma_v^2$.  From the 
results of \refsec{vir}, we expect that when averaged over many halos,
\be
\frac{\s_{v,\rm MG}^2}{\s_{v,\rm GR}^2} = \gvir,
\label{eq:sratio}
\ee
where $\s_{v,\rm MG}^2$ is the velocity dispersion measured in the
modified gravity simulations, while $\s_{v,\rm GR}^2$ is
measured in the corresponding GR simulations.  Note that
in this measurement, we can only compare the average of many halos
in the modified gravity simulations to that in GR, rather than calculating 
$\g$ on a halo-by-halo basis.  Hence, \refeq{sratio} results in a 
noisier measurement of $\gvir$ than \refeq{gvirsim}.  However,
the particle velocity dispersion, which has gone through the relaxation and 
virialization process, is much more closely related to observables than the 
averaged gravitational force strength \refeq{gvirsim}, which can never be 
measured directly in reality.  
Thus, it is worthwhile to cross-check our results obtained from \refeq{gvirsim}
with the halo velocity dispersions.
% !!!!!!!!!!!!!!!!!!!!!!!!!!!
\begin{figure}[t!]
\centering
\includegraphics[width=0.49\textwidth]{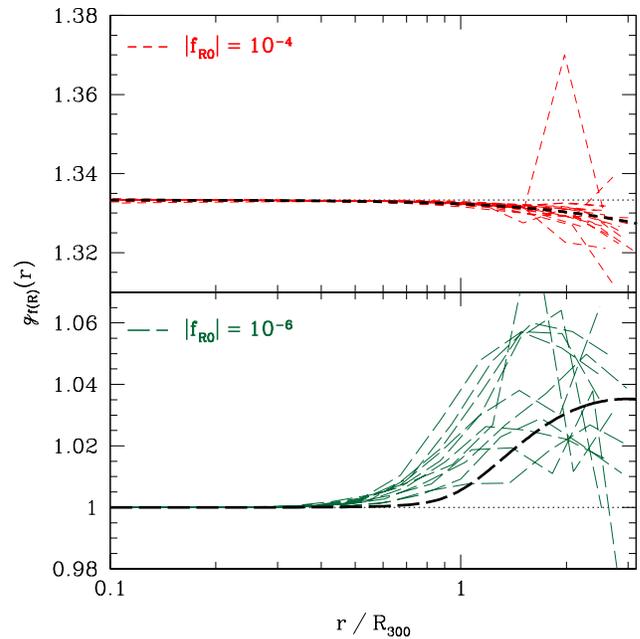}
\caption{$\g_{f(R)}(r)$ measured using \refeqs{Wofr}{gsim} (thin lines), for the most 
well-resolved halos ($R_{300}> 10$ grid cells) in the $f(R)$ simulations, for
the strong field $|f_{R0}|=10^{-4}$ and weak field $|f_{R0}|=10^{-6}$.  The
mass range of the halos shown here is $M_{300}=1.6-7\times10^{14}\Msunh$.  The
thick lines show the results of the relaxation code (for $M_{300}=3\times 10^{14}\Msunh$). 
\label{fig:Wprof-fR-4-6}}
\end{figure}
% !!!!!!!!!!!!!!!!!!!!!!!!!!!

In order to determine for which halos we can reliably measure \refeq{gvirsim}
and \refeq{sigmav},
we calculate the velocity dispersion of halos in the standard GR simulations
and compare it to the expected virial scaling.  
\reffig{sigmav_res}
shows the measured velocity dispersion scaled as $\sigma_v^2 / \Mv^{2/3}$, as 
a function of the halo radius $R_{300}$ in grid cells, for the different 
simulation boxes.  Since the virial theorem
has been found to hold in simulated halos \cite{EvrardEtal}, 
$\sigma_v^2 / \Mv^{2/3}$ should be independent of the halo mass (\refsec{th}).
We see that, within the significant scatter, this is approximately true
for halos that are sufficiently resolved.  We thus 
place a radius cut of $\Rv \geq 5.4$~grid cells.  Since 
$\Mv = (4\pi/3)\:\D\,\rhob\,\Rv^3$, this corresponds
to a fixed mass cut for a given simulation box size, which is listed
in \reftab{param}.  Fortunately, the statistics are then sufficient to study
$\gvir$ and $\s_v^2$ as functions of mass.  

%%%%%%%%%%%%%%%%%%%%%%%%%%%%%%%%%%%%%%%%%%%%%%%%%%%%%%%%%%%%%%%%%%%%%%%%%%%%%
%\subsection{$f(R)$}
\subsection{\textbf{\textit{f(R)}}}
\label{sec:fRsim}

We begin with the measurement of $\g_{f(R)}$ for the well-resolved halos.  
\reffig{Wprof-fR-4-6} shows the simulation measurements and
predictions of the spherical relaxation code, for the strong field 
($|f_{R0}|=10^{-4}$) and the weak field ($10^{-6}$).  As expected from
\reffig{gvirfR}, the halos in this mass range ($M_{300}\approx 1.6-7\times10^{14}\Msunh$) are unscreened in the strong field simulations,
$\g_{f(R)}=4/3$, and screened in the weak field, $\g_{f(R)}\rightarrow 1$.  
In the latter case, there is a regime
around 1--3 $R_{300}$ where the screening is not complete.  At larger
distances, the Yukawa suppression again becomes noticeable 
($\l_C \approx 3$~Mpc for $|f_{R0}|=10^{-6}$).  The numerical results for
the spherically symmetric case match the overall behavior well for both
field values, although there is a hint that it slightly overestimates the
screening in the weak field case.

% !!!!!!!!!!!!!!!!!!!!!!!!!!!
\begin{figure}[t!]
\centering
\includegraphics[width=0.49\textwidth]{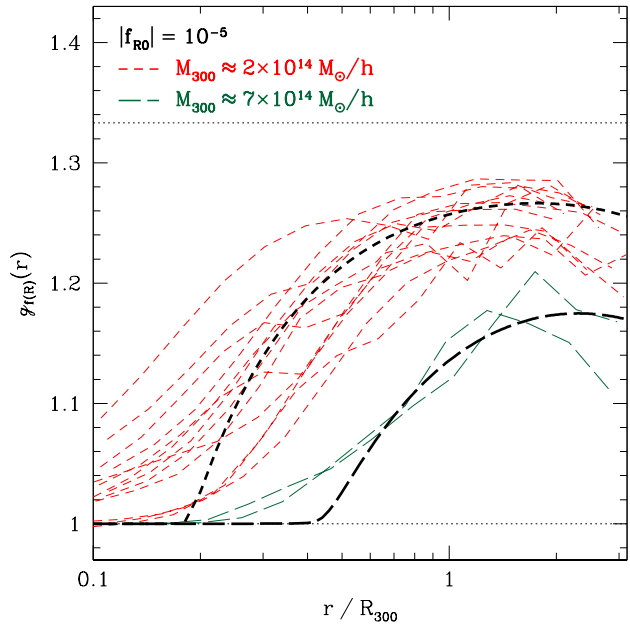}
\caption{Same as \reffig{Wprof-fR-4-6}, but for the intermediate field
value $|f_{R0}|=10^{-5}$.  We have separated the halo sample into lower
mass halos with $M_{300} = 1.6-2.5\times10^{14}\Msunh$, and two higher
mass halos with $M_{300}=6-7\times10^{14}\Msunh$.
\label{fig:Wprof-fR}}
\end{figure}
% !!!!!!!!!!!!!!!!!!!!!!!!!!!

% !!!!!!!!!!!!!!!!!!!!!!!!!!!
\begin{figure}[t!]
\centering
\includegraphics[width=0.49\textwidth]{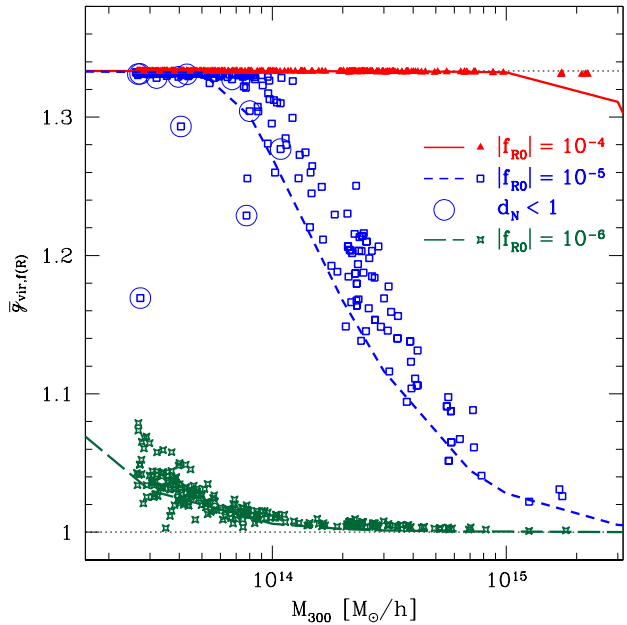}
\caption{$\gvir$ measured via \refeq{gvirsim} for well-resolved halos 
($R_{300}>5.4$~grid cells) in 
the $f(R)$ simulations (points).  The results confirm the theoretical 
predictions from \refsec{fR}, shown as lines (spherical relaxation results
from \reffig{gvirfR}).  The circled points are halos which have a more massive
halo in their immediate vicinity (see text). 
\label{fig:gvirsim_fR}}
\end{figure}
% !!!!!!!!!!!!!!!!!!!!!!!!!!!
The results for the intermediate field value $|f_{R0}|=10^{-5}$ are shown in
\reffig{Wprof-fR}; this case is most interesting since the few$\times 10^{14}\Msunh$
halos are in the transition region between the screened and unscreened 
limits (\reffig{gvirfR}).  Hence we have split the halo sample into a
lower mass sample around $2\times 10^{14}\Msunh$ and a high mass sample
with two halos around $7\times 10^{14}\Msunh$.  Clearly, the scatter in the
modified force profiles is significant.  Nevertheless, the stronger
screening effect in the higher mass halos is noticeable.  The spherical relaxation
results (\refsec{fR}), which were calculated separately for the mean halo mass of each 
sample, match the full simulation profiles remarkably well.  At small radii,
the transition to the fully screened values is apparently too steep.  A
possible explanation for this is that the halos in the N-body simulations
are not truly spherical, but in general triaxial.  A triaxial halo will
have a somewhat shallower potential well, reducing the chameleon screening
effect.  Furthermore, the screening will happen at different radii along
the different axes, so that a potentially sharp transition in the spherical
case is washed out over a certain radius range.  In addition, the innermost
Newtonian potential well is not as deep in the simulations as predicted
for a perfect NFW halo due to the finite resolution.  We also reiterate that
the profiles only become reliable at $r \sim 0.3-0.4\:R_{300}$ for these
halos.

Next, we look at $\gvirf{\rm meas}$ in the larger sample
of halos.  \reffig{gvirsim_fR} shows the results for the three field
values, and the predictions of $\gvirf{f(R)}(M)$ from the spherical
relaxation code.  We again see a very good match for all field 
values and over the entire mass range probed by this halo sample,
$3\times10^{13}\Msunh < M_{300} < 3\times 10^{15}\Msunh$.  For the
intermediate field value, which again shows the most interesting behavior
in this mass range, we see that the screening effect is slightly overpredicted
in the spherically symmetric approximation.  Again, this could be due to
halo triaxiality and to resolution effects which reduce the value of $\Psi_N$
in the inner parts of the halo.

For the intermediate field, some outliers are seen in \reffig{gvirsim_fR}.  
These halos, especially around $3-8\times10^{13}\Msunh$ show a stronger screening of
the modified forces than the large majority of halos at that mass, and 
stronger than predicted for isolated spherical NFW halos.  This would seem consistent
with halos being screened by a larger scale potential well in which
they are situated.  To test this hypothesis, we have selected halos
in the intermediate field simulations which have a more massive neighboring 
halo in their immediate vicinity.  More
precisely, we ask that
\be
d_N \equiv \frac{|\vx_{N}-\vx_{h}|}{R_{\D,N}+R_{\D,h}} < 1,
\ee
where $h$ denotes the halo itself and $N$ denotes the closest neighboring halo\footnote{This
criterion formally says that the halos are overlapping. Such an overlap is 
unavoidable when defining halos via spherical overdensities.  In our halo
finding algorithm, the particles in the overlap region are not double-counted,
but counted towards the more massive halo.} with 
$M_{\D,N} > M_{\D,h}$.  These halos make up less than
5\% of the whole sample and are circled in \reffig{gvirsim_fR}.  In fact, three
of the outliers have a close massive neighbor, which is strong evidence
for the hypothesis of environmental effects as cause for the enhanced screening 
(the fourth most obvious outlier has $d_N\approx 1.2$).  

% !!!!!!!!!!!!!!!!!!!!!!!!!!!
\begin{figure}[t!]
\centering
\includegraphics[width=0.49\textwidth]{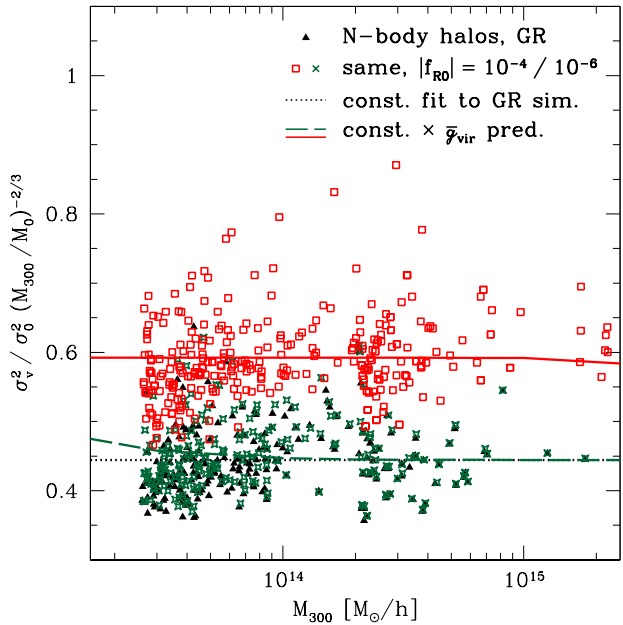}
\caption{Scaled velocity dispersion $\sigma_v^2/M^{2/3}$ measured in GR and
$f(R)$ simulations.  The measurements were scaled to values expected for NFW 
halos (\refsec{th}): 
$\sigma_0^2 = 1.79\times10^{-5} c^2$, $M_0=10^{15}\Msunh$.  The dotted black
line shows a constant fit to the GR results.  Solid and dashed lines
show the predictions of the model of \refsec{fR} scaled by the GR value.
\label{fig:sigmav_fR4}}
\end{figure}
% !!!!!!!!!!!!!!!!!!!!!!!!!!!

% !!!!!!!!!!!!!!!!!!!!!!!!!!!
\begin{figure}[t!]
\centering
\includegraphics[width=0.49\textwidth]{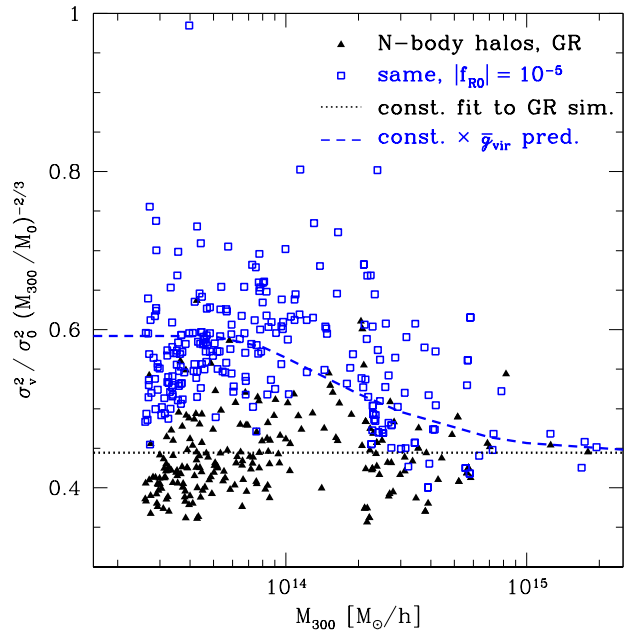}
\caption{Same as \reffig{sigmav_fR4}, but for the intermediate $f(R)$ field
value $|f_{R0}|=10^{-5}$.
\label{fig:sigmav_fR5}}
\end{figure}
% !!!!!!!!!!!!!!!!!!!!!!!!!!!

Finally, we can look at the effect of the modified forces on the particle
velocity dispersion of halos, a noisier measurement but one that probes
the effect after the reprocessing through gravitational collapse and virialization.  
\reffig{sigmav_fR4} shows the scaled velocity dispersion $\sigma_v^2/M^{2/3}$, 
for the same halos as in \reffig{gvirsim_fR} and scaled to values expected 
from \refsec{th}, as a function of mass.  We show the results for
GR simulations as well as the weak ($|f_{R0}|=10^{-6}$) and strong field 
($|f_{R0}|=10^{-4}$) $f(R)$ cases.  After fitting a constant to the 
GR simulations, we multiply the theoretical predictions by this
constant.  
Albeit noisy, the results of the halo-by-halo measurement of $\gvirf{f(R)}$
are confirmed:  for the strong field, all halos are
in the linearized field regime where forces, and hence $\sigma_v^2$ are enhanced
by a factor of 4/3.  For the weak field, all halos except at the very lowest
masses probed by the simulations are in the chameleon regime.  In case
of the intermediate field ($|f_{R0}|=10^{-5}$), \reffig{sigmav_fR5} shows that the transition 
between screened and unscreened regimes at few$\times 10^{14}\Msunh$ 
is indeed seen in the halo velocity dispersions as well.  
These results confirm that the theoretical predictions for the modified
gravitational force can in principle be probed by observable quantities
such as velocity dispersions (see \refsec{obs}).

%%%%%%%%%%%%%%%%%%%%%%%%%%%%%%%%%%%%%%%%%%%%%%%%%%%%%%%%%%%%%%%%%%%%%%%%%%%%%
\subsection{DGP}
\label{sec:DGPsim}

The force modifications in DGP are, to first order, independent of the halo mass
and environment.  However, they do depend on the detailed halo profile.  
Before comparing the predictions from \refsec{DGP} with the simulation 
results, we have to take into account the effects of the finite resolution.  
While the NFW profile we used throughout \refsec{th} is a very good 
match to high-resolution simulations, in our fixed-grid simulations
the density profile is in fact softened on scales of a grid cell.  This 
softening of the density profile
will affect $\g_{\rm DGP}(r)$ through the average overdensity within $r$.  Thus,
for comparison with the simulation results we assume a ``capped'' density 
profile instead of NFW all the way to $r=0$.  More precisely, we cap the 
density profile at a constant value of $\rho_{\rm cap}=\rho_{\rm NFW}(r_{\rm cap})$ 
for $r < r_{\rm cap}$\footnote{Note that the halo radius for a given mass is 
slightly increased when using the capped density profile, in order to match
the fixed overdensity $\D=200$.}.  For the halos measured in the smallest box, a natural
choice is $r_{\rm cap}=r_{\rm cell}=0.125\Mpch$ (\reftab{boxes}).  \reffig{gDGP} shows
the effect of this softened density profile on $\g_{\rm DGP}(r)$. In particular,
it increases the force modification since the inner density is suppressed, 
thus artificially weakening the Vainshtein mechanism.  

% !!!!!!!!!!!!!!!!!!!!!!!!!!!
\begin{figure}[t!]
\centering
\includegraphics[width=0.49\textwidth]{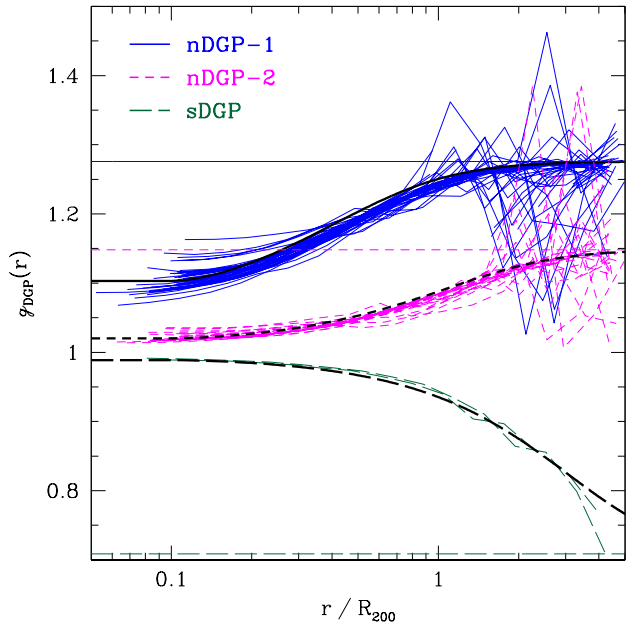}
\caption{$\g_{\rm DGP}(r)$ measured using \refeqs{Wofr}{gsim} for the most 
well-resolved halos ($R_{200}> 10$ grid cells) in the DGP simulations (thin
lines).  The thick lines show the prediction of \refeq{gDGP}, using a
capped NFW profile with $r_{\rm cap}=0.125\Mpch$ (see text).  The thin horizontal lines show
$\g_{\rm DGP,lin}$ for each model.  The halos shown here
have masses $M_{200}=1.6-7\times10^{14}\Msunh$.
\label{fig:Wprof-DGP}}
\end{figure}
% !!!!!!!!!!!!!!!!!!!!!!!!!!!

\reffig{Wprof-DGP} shows the measured $\g_{\rm DGP}(r)$ from the simulations,
together with the predictions using the capped density profile.  First,
it is evident that the scatter in $\g$ is much smaller in DGP than it is
for $f(R)$, due to the locality of the Vainshtein mechanism.  For all three
models, the
agreement of the simple spherically symmetric NFW model with the simulations
is impressive.  Note that we have not adjusted any parameters to match
the simulation results; this measurement thus also constitutes a nontrivial test 
of the DGP simulations. 
  At $r\sim R_{200}$, the theoretical prediction slightly underestimates the
suppression of the force modification (by $1-3$\%), which is presumably due to slight
differences in the actual density profiles from the one assumed in the 
predictions (pure spherical NFW profile).  The large scatter at $r \gtrsim 2 R_{200}$ is due to
the effect of gravitationally unbound ambient matter in the environment of the halos, which dominates
$\d\rho$ at these distances.  Note that in particular for nDGP--1, the 
Vainshtein mechanism does not completely suppress the force
modifications  within halos even on scales as small as $\sim 100$~kpc.

% !!!!!!!!!!!!!!!!!!!!!!!!!!!
\begin{figure}[t!]
\centering
\includegraphics[width=0.49\textwidth]{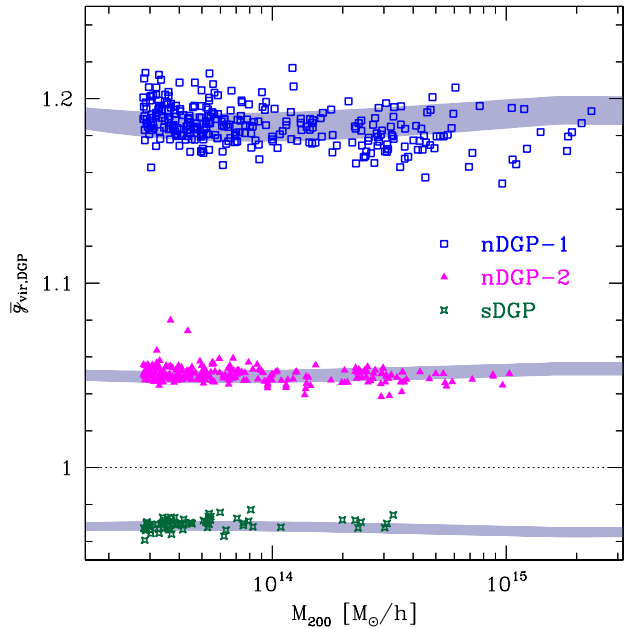}
\caption{$\gvirf{\rm DGP}$ measured using \refeq{gvirsim} for well-resolved 
halos ($R_{200} > 5.4$~grid cells) in the DGP simulations (points).  The
shaded bands show the model predictions from \refsec{DGP} with a variation
in the halo concentration by $\pm 20$\%.  We assumed capped NFW profiles with
$r_{\rm cap}=0.125\Mpch$.
\label{fig:gvirsim_DGP}}
\end{figure}
% !!!!!!!!!!!!!!!!!!!!!!!!!!!

% !!!!!!!!!!!!!!!!!!!!!!!!!!!
\begin{figure}[t!]
\centering
\includegraphics[width=0.49\textwidth]{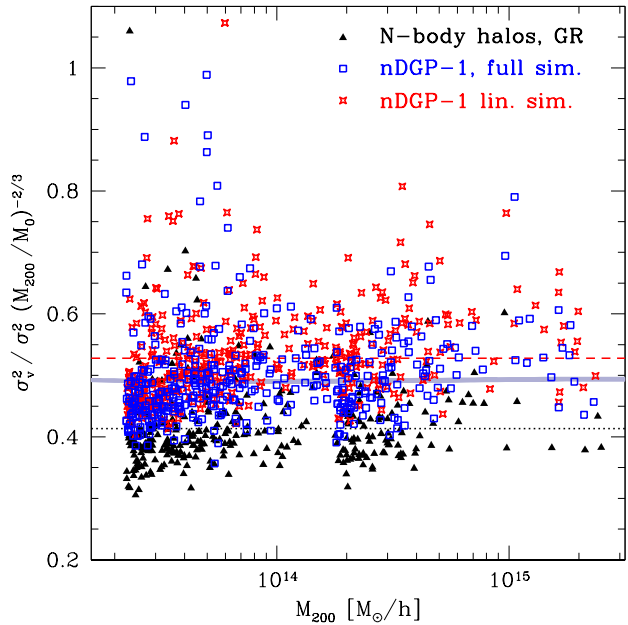}
\caption{Scaled velocity dispersion $\sigma_v^2/M^{2/3}$ measured in GR and 
nDGP--1 simulations ($\sigma_0$ and $M_0$ are as defined in 
\reffig{sigmav_fR4}).  The black dotted line shows the average value measured 
for the GR simulations.    The red line shows this value multiplied by 
$\gvirf{\rm DGPlin}=1+1/(3\beta)$.  The shaded band shows the corresponding 
prediction for $\gvirf{\rm DGP}$ from \reffig{gvirsim_DGP}.  
\label{fig:sigmav_nDGP2}}
\end{figure}
% !!!!!!!!!!!!!!!!!!!!!!!!!!!

We now turn to $\gvir$ as measured from \refeq{gvirsim} in the halo sample with
$R_{200} \geq 5.4$~grid cells.  \reffig{gvirsim_DGP} shows the measurements for
the three DGP models.  As expected, $\gvirf{\rm DGP}$ is approximately 
constant as a function of mass.  The model predictions from \refsec{DGP} are
shown as gray bands.  Here, we have used the concentration relation 
\refeq{cM} (more precisely, $c=\max\{4,c(M)\}$), and the width of the band
reflects a $\pm 20$\% spread in concentration.  We again assumed a capped
NFW profile with $r_{\rm cap}=0.125\Mpch$, the effects of which are noticeable as a
slight increasing trend of $\gvir(M)$ in going towards the low-mass end for nDGP--1.  Note
that here we have included halos from different simulation box sizes 
$\Lbox=64-256\Mpch$, though the majority comes from the smallest box.  Hence,
one might wonder whether different values of $r_{\rm cap}$ are required
for different box sizes.  However, within the limited statistics the
measurements of $\gvir$ from halos in different box sizes are in agreement.  
Hence, the data do not seem to require such a correction.  We conclude that,
within the uncertainties due to the halo density profiles, the measured values
of $\gvirf{\rm DGP}$ are entirely consistent with the predictions.  
Furthermore, the scatter in the measured $\gvir$ appears consistent with 
that expected for intrinsic variations of halo density profiles
($\D c/c \sim 0.2$).  

The results for $\gvirf{\rm DGP}$ are confirmed by the particle velocity
dispersions of halos.  
\reffig{sigmav_nDGP2} shows the scaled velocity dispersion $\s_v^2/M^{2/3}$ in
nDGP--1 and the corresponding GR simulations.  For comparison,
we also show the result for the {\it linearized} DGP simulations, which
use the scale-independent (but redshift-dependent) force enhancement obtained
when linearizing the DGP equations \cite{DGPMpaper,DGPMpaperII}.  
Within the significant scatter in the $\s_v$ measurement, we found no 
significant evolution of the force enhancement with mass, as expected given
the small trends with mass in \reffig{gvirsim_DGP}.

In order to quantitatively compare the simulation results with model
predictions, we determined the mean of $\sigma_v^2 / M^{2/3}$ for
each simulation type.  In each case, the error on this mean is obtained
by dividing the RMS scatter by $\sqrt{N_{\rm halos}}$.  The measured ratio of
the scaled velocity dispersion in the DGP simulations to that in the GR 
simulations is found to be
\be
\gvir(\mbox{full DGP, meas}) = 1.212 \pm 0.014.
\ee
This is indeed close to the range of the theoretical predictions (for 
a capped NFW profile), $1.18-1.2$ (\reffig{gvirsim_DGP}).  As expected, the 
ratio measured in the linearized DGP simulations, 
$\gvir(\mbox{lin. DGP, meas}) = 1.288 \pm 0.014$ is 
in excellent agreement with the predicted value of $1 + 1 / (3\beta) = 1.276$. 
  Similar conclusions
hold for the velocity dispersions measured in the sDGP and nDGP--2 
simulations, although the results are less constraining due to the smaller
force modifications $|\gvir-1|$ in those models.

%%%%%%%%%%%%%%%%%%%%%%%%%%%%%%%%%%%%%%%%%%%%%%%%%%%%%%%%%%%%%%%%%%%%%%%%%%%%%
\section{Application to observations}
\label{sec:obs}

Observables linked to dynamical masses can be broadly classified into
two categories.  First, one can measure the velocity distribution
of collisionless ``tracer particles'', such as galaxies within galaxy clusters,
or stars within galaxies.  For a dynamically relaxed system, the
kinetic energy $T$ inferred from the velocity distribution is proportional
to the potential energy $W$ (\refsec{vir}), which can be converted into a 
mass estimate $\Mdyn$ (we again assume
a mass definition in terms of an average interior density $\rhob\D$).  Several
assumptions have to be made in order to obtain the mass estimate.  First,
one has to assume the galaxies or stars are unbiased tracers of the full
matter velocity field (including dark matter).  Since member galaxies of
a cluster generally reside in overdense substructure (subhalos) of the cluster
halo, their velocities might differ systematically from that of the overall
matter.  Simulation studies \cite{FalDiemand,LauEtal} have shown that this 
velocity bias is expected to be on the order of $\sim 10$\% or less, depending
on how galaxies are selected.  Further, one has to make assumptions about
the density profile shape, and the anisotropy of the velocity distribution,
since only the line-of-sight component of the velocity is observed.  
Nevertheless, our idealized measurement of the dark matter velocity dispersion in the 
simulations shows that at least in principle, $\sigma_v^2$ is indeed a good 
tracer of the modified force $\gvir$. 

Another set of observations linked to dynamical masses is measurements
of the hot ionized gas in galaxy clusters.  One technique is to detect the
thermal bremsstrahlung in X-rays; another is to measure the upscattering
of CMB photons off the hot electrons via the Sunyaev-Zeldovich (SZ) effect.  
In both techniques, one measures
a line-of-sight integral of the electron pressure, with an additional
weighting by the electron density in the case of X-rays (since the rate
of bremsstrahlung emission is $\propto n_e\:n_p = n_e^2$).  
With some assumptions on the
density profile for the baryons, X-ray and SZ signals can be converted 
into a measurement of the electron pressure as function of $r$. 
Instead of the virial theorem which holds for a collisionless system,
we now use hydrostatic equilibrium which is a good assumption at least
for dynamically relaxed systems:
\be
\frac{d P}{dr} = \rho_{\rm gas} \frac{d\Psi}{dr},
\label{eq:hydeq}
\ee
where $P$ is the total pressure and $\rho_{\rm gas}$ is the mass 
density of the gas, respectively.  The difficulty observationally is in
measuring the left-hand side of \refeq{hydeq}: only the thermal contribution
to $P$, $P_{\rm therm} \sim n_{\rm gas} k T$ is directly
measurable, while non-thermal contributions from e.g. cosmic rays,
bulk flows, and magnetic fields are much harder to estimate.  Nevertheless,
with appropriate systematic error bars, \refeq{hydeq} is a probe of the 
gravitational force $d\Psi/dr$.

In summary, a variety of observations lead to estimates of certain weighted 
averages of the gravitational force,
\be
W_{\rm obs} = \int d^3\vx\:\rho_{\rm obs}(\vx)\:\vx\cdot\vn\Psi(\vx),
\ee
where $\rho_{\rm obs}$ is an effective weight function.  In case of
X-ray and SZ measurements, it is related to $\rho_{\rm gas}^2$ and
$\rho_{\rm gas}$, respectively, but will be modified by instrumental effects
such as the limited instrument aperture.  Similarly, for galaxy velocity 
dispersions
in clusters, $\rho_{\rm obs}$ is proportional to the number
density of observed galaxies (again, with observational weights and boundaries
folded in).  

Now we can use \refeq{Emod} together with the fact that 
$\Psi \propto \Mv^{2/3}$, so that $W \propto M^{5/3}$ [\refeq{W}].  Then, 
if the observational mass estimate is done based on ordinary gravity, so
that in GR the mass estimate equals the true mass $\Mv$, the 
resulting mass estimate $\Mdyn$ in modified gravity is in fact
\be
\Mdyn = \gbar_{\rm obs}^{3/5}\:\Mv.
\label{eq:Mdyn}
\ee
Here $\gbar_{\rm obs}$ is a weighted integral using \refeq{wvir}, with
$\rho$ replaced by the effective weight $\rho_{\rm obs}$.  Note that in
general $\gbar_{\rm obs}$ will depend on the true mass $\Mv$ itself.  

Since the true mass can in principle
be obtained from weak or strong lensing, a comparison of lensing mass with
the dynamical mass \refeq{Mdyn} can be used to measure the modified
forces in $f(R)$ or DGP.  Again, it is important to take into account
the unavoidable observational weighting that is being done in the measurements of 
both $\Mdyn$ and $\Mv$.  

Recently, the SLACS sample of elliptical galaxies acting as strong lenses 
\cite{SLACS} has been used
to constrain deviations from GR \cite{Smith,SchwabEtal09}.  Furthermore,
using a similar argument as the thin shell condition \refeq{thinshell}
(\refsec{fR}), \cite{Smith} has shown that these measurements constrain 
the $f(R)$ model considered here at the level of $|f_{R0}|\lesssim 2\times
10^{-6}$.  For these constraints one has to make some assumptions on the
potential well of the lens galaxy, for example that it is dominated by the 
density distribution of the inner few kpc, thus neglecting any larger scale 
potential well.  
% (which could be present, e.g., if the galaxy was a member of a massive 
%cluster).  
As we have seen, the magnitude of the force modification 
in $f(R)$ can depend somewhat on the environment.  In
particular, we found that a subset of halos around $3-8\times 10^{13}\Msunh$
(at the low mass end of the range accessible to the simulations) is
screened much more strongly than expected for isolated halos, consistent
with an effect of the large scale environment.  Nevertheless,
strong lens galaxies offer a quite powerful probe of gravity on kpc scales,
if the environmental effects can be understood.  

On larger scales, the comparison of dynamical and lensing 
masses of massive galaxy clusters can be interesting since they dominate
their local environment, so that environmental effects should be negligible.  
  Also, for cluster-scale masses we were able to validate 
our theoretical models for $\gvir$ directly with the modified gravity simulations 
(\refsec{sim}).  However, for clusters it is preferable to measure the dynamics and
lensing at large scales: first, the deviations from GR quickly shrink close
to the cluster core owing to the chameleon and Vainshtein mechanisms;  second,
baryonic effects on the observables and the density profile, such as cooling 
and AGN feedback, are expected to be
less significant at greater distances from the cluster center.  

It is also possible to use dynamic mass estimates of clusters by themselves,
without direct comparison to lensing masses.  As shown in \cite{WhiteEtal93,EkeEtal98,BorganiEtal01,Vikhlinin,fRcluster,MantzEtal09}, the 
abundance of massive clusters is a sensitive probe of the growth
of structure as well as gravity.  When comparing the observed
cluster mass function measured using a {\it dynamical} mass measure with 
modified gravity predictions, it is necessary to take into account the
effect of the modified forces on the mass estimates as well.  In order
to estimate the effect on the observed mass function, we use
\refeq{Mdyn}, setting $\gbar_{\rm obs} = \gvir$, the idealized quantity we 
have modeled and calibrated with simulations.  Dynamical mass measures
(i.e. velocity dispersions) in our simulations are noisy (\refsec{sim}),
thus we have simply rescaled the mass of each halo in the modified gravity
simulations by our theoretical model of $\gvir(M)$ for the given cosmology.  
% !!!!!!!!!!!!!!!!!!!!!!!!!!!
\begin{figure}[t!]
\centering
\includegraphics[width=0.49\textwidth]{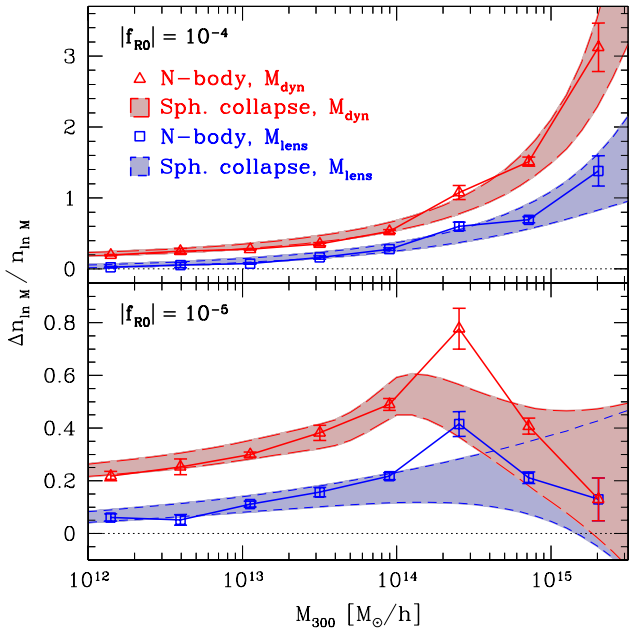}
\caption{Mass function enhancement in $f(R)$ relative to a $\L$CDM cosmology
with the same expansion history, 
$n_{\ln M}(f(R))\,/\,n_{\ln M}(\Lambda\mbox{CDM})-1$,
for $|f_{R0}|=10^{-4}$ (\textit{top panel})
and $10^{-5}$ (\textit{bottom panel}).  The points show simulation results, while
the shaded bands show spherical collapse predictions \cite{HPMhalopaper} (see text).  
Results are shown for the mass function $n_{\ln \Mv}$ in terms of the lensing
mass, and $n_{\ln \Mdyn}$ in terms of the dynamical mass (see text; $\D=300$ in both cases).
\label{fig:dndm_fR}}
\end{figure}

\reffig{dndm_fR} shows the relative enhancement of the mass function in 
$f(R)$ gravity with respect to $\L$CDM, when measured using lensing
masses (i.e. ``true'' $M_{300}$) and dynamical masses, for\footnote{Since 
all halos above $\sim 10^{13}\Msunh$ are screened for the
small field $|f_{R0}|=10^{-6}$, the dynamical mass function is essentially
equal to the lensing mass function for most of the mass range, and is not
repeated here (see \cite{HPMhalopaper}).} $|f_{R0}|=10^{-4}$
and $10^{-5}$. Clearly, the observed abundance of halos is further enhanced 
when measured in terms of dynamical masses.  In the mass range where halos
are unscreened, the mass function enhancement is boosted by a factor of two
or more.  This is because the
dynamical mass estimate is a factor of $(4/3)^{3/5}\approx 1.19$ higher than
the lensing mass in $f(R)$ gravity in the unscreened case, in conjunction with
the steeply falling mass function.  Constraints on $f(R)$ gravity
from X-ray clusters could thus be significantly improved
by using the dynamical mass function instead of the true or lensing mass function
which was used in \cite{fRcluster}.  Note the sharp turnover in the 
mass function enhancement for the intermediate field value.  This transition due
to the onset of the chameleon mechanism is already present in the lensing
mass function \cite{HPMhalopaper}.  Since $\gvir$ transitions from 4/3 to
1 in this mass range as well, the effect is enhanced in the dynamical mass
function.  

% !!!!!!!!!!!!!!!!!!!!!!!!!!!
\begin{figure}[t!]
\centering
\includegraphics[width=0.49\textwidth]{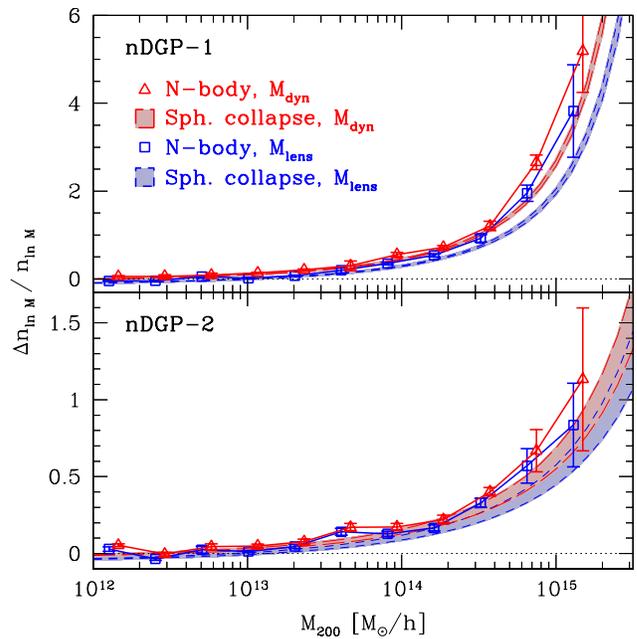}
\caption{Same as \reffig{dndm_fR}, but for the DGP models nDGP--1
(\textit{top panel}) and nDGP--2 (\textit{bottom panel}).  The simulation results
for the dynamical mass (red triangles) have been displaced horizontally for clarity.  The shaded
band shows the spherical collapse model of \cite{DGPhalopaper}.
\label{fig:dndm_DGP}}
\end{figure}
% !!!!!!!!!!!!!!!!!!!!!!!!!!!

The shaded bands in \reffig{dndm_fR}
show the spherical collapse predictions presented in \cite{HPMhalopaper}.  
These are based on the linear $f(R)$ matter power spectrum together with the 
Sheth-Tormen prescription, using two sets of collapse parameters derived
for limiting cases of spherical collapse in $f(R)$ (enhanced forces throughout,
and unmodified forces).  We rescaled the predictions in terms of lensing
mass given in \cite{HPMhalopaper} to the dynamical mass via
\be
n_{\ln\Mdyn} \equiv \frac{dn}{d\ln\Mdyn} = \frac{d\ln\Mv\quad\;\;}{d\ln\Mdyn}\: n_{\ln\Mv}.
\ee
As expected, the predictions in terms of dynamical mass perform equally
well as those for the lensing mass.  Since our prediction for $\gvir$
includes the chameleon mechanism, the predictions for the intermediate field
value show a corresponding transition at approximately the right mass.  Still,
the predictions do not match the simulation results completely due to the
shortcomings of our simple spherical collapse model \cite{HPMhalopaper}.  

\reffig{dndm_DGP} shows the corresponding results for the two normal-branch
DGP models nDGP--1 and nDGP--2.  The effect is less dramatic on the DGP
mass function, since $\gvir$ in DGP is generally smaller than in $f(R)$.  
Nevertheless, the impact especially for nDGP--1 is significant, an additional
implying an abundance boost of $\sim$50\% at high masses.  The shaded bands in \reffig{dndm_DGP}
again show a spherical collapse model \cite{DGPhalopaper}, which uses the
analytical solution for the modified forces in DGP in the spherically
symmetric case as one limiting case of spherical collapse in DGP.  The other 
limit is given by using the linearized expression for the modified forces.  
Again, the spherical collapse model performs equally well for the 
mass function in terms of dynamical mass as for the lensing mass function.

%%%%%%%%%%%%%%%%%%%%%%%%%%%%%%%%%%%%%%%%%%%%%%%%%%%%%%%%%%%%%%%%%%%%%%%%%%%%%
\section{Conclusions}
\label{sec:concl}

In this paper, we have studied the dynamics of matter within bound cosmic
structures, i.e. dark matter halos, in $f(R)$ and DGP.  The
potential governing matter dynamics can differ from the lensing potential
by $20-30$\% in these models.  These unique signatures of modified gravity
can be observed by comparing dynamical and lensing mass estimates of
clusters or galaxies.  Furthermore, they strongly influence the observed 
abundance
of massive clusters when measured via dynamical mass proxies such as 
X-rays or the SZ effect.  For example, the enhancement of the cluster
abundance in $f(R)$ (with respect to $\L$CDM) at a fixed {\it dynamical} mass 
can be roughly twice that measured if the mass is based on lensing measurements.   
These signatures in the dynamics are also relevant for large-scale structure
observations, such as the redshift-space power spectrum or correlation function
on small scales.

However, since halos are
highly non-linear objects, the peculiar chameleon and Vainshtein mechanisms
play a crucial role, as they are necessary in order to restore General Relativity in 
high-density environments.  Thus, the dynamics in these
models can only be rigorously studied through N-body simulations which
include the non-linear mechanisms of $f(R)$ and DGP consistently.  

In the case of $f(R)$, 
the chameleon mechanism is triggered once the depth of the potential
well is comparable to the background value of the scalar field.  The
suppression of the force modifications within a halo thus depends not only 
on the halo mass but also its environment.  Consequently, we found
significant scatter from halo to halo in the force modification $\g$ measured
in the $f(R)$ simulations.  Furthermore, we identified a subset of halos
which are in the close vicinity of massive neighbors, and which show
a much stronger suppression of the force modifications than expected for 
isolated halos.  In the majority of cases however, the simulation results confirm the
basic expectation that halos are ``unscreened'' below a certain threshold
mass determined by the potential well and the field value, whereas GR 
is restored at higher masses.  Furthermore, a 
simple model based on the
spherically symmetric solution of the field equations provides
a good match to the scale- as well as mass-dependence of the force
modifications in $f(R)$.

In DGP, the non-linear suppression of the force modifications through
the Vainshtein mechanism is much less dependent on halo mass and details
of the large scale environment.  Instead, the crucial quantity is
the average mass density within a given radius.  Thus, uncertainties in 
the semi-analytic predictions for DGP are mainly due to the density profile,
and are already quite small.  When taking into account the force resolution
of the simulations, our predictions provide an excellent fit to the simulation
measurements.  Since the basic assumptions of the model,
in particular spherical symmetry seem to hold well, we expect that force modifications 
can be predicted very accurately in DGP, provided the density profile is known
sufficiently well.

Given that our semi-analytic models appear to capture the mass- and 
scale-dependence of the modified forces correctly for both $f(R)$ and DGP, 
they can be useful in
extending predictions beyond the limits of resolution and parameter space of 
the simulations.  This will be necessary in particular for the comparison 
with observations.  

While this study is specific to $f(R)$ and DGP, it shows the qualitative
features expected in observations of dynamics from viable modified gravity 
models, which employ a non-linear mechanism to restore GR locally.  In the
outer regions of massive clusters, as well as in lower mass objects, these models generally predict
order unity deviations from GR.  Observations in this regime thus offer the 
perspective of closing
the last remaining loopholes for significant modifications to gravity on
large scales.

\vspace*{1cm}
\begin{acknowledgments}
I would like to thank Wayne Hu, Tristan Smith, Mark~Wyman and Donghai~Zhao 
for discussions and comments on the paper.

This work was supported by the Gordon and Betty Moore Foundation at Caltech.  
The simulations used in this work have been performed on the Joint 
Fermilab - KICP Supercomputing Cluster, supported by grants from Fermilab,
Kavli Institute for Cosmological Physics, and the University of Chicago. 

\end{acknowledgments}

%\clearpage
%%%%%%%%%%%%%%%%%%%%%%%%%%%%%%%%%%%%%%%%%%%%%%%%%%%%%%%%%%%%%%%%%%%%%%%%%%%%%
%%%%%%%%%%%%%%%%%%%%%%%%%%%%%%%%%%%%%%%%%%%%%%%%%%%%%%%%%%%%%%%%%%%%%%%%%%%%%
%\bibliographystyle{arxiv_physrev}
\bibliography{DGPM}

\end{document}